\documentclass[aps,prc,twocolumn,superscriptaddress,showpacs,nofootinbib]{revtex4-1}
\usepackage[dvips]{graphicx}
\usepackage{amsmath,amssymb}
\usepackage{bm}
\usepackage[T1]{fontenc}
\usepackage{sidecap}

\begin{document}
\title{Thermal quasiparticle random-phase approximation calculations of
    stellar electron capture rates with the Skyrme effective interaction}

\author{Alan~A.~Dzhioev}
\email{dzhioev@theor.jinr.ru}
\affiliation{Bogoliubov Laboratory of Theoretical Physics, JINR, 141980, Dubna, Russia}
\author{A.~I.~Vdovin}
\email{vdovin@theor.jinr.ru}
\affiliation{Bogoliubov Laboratory of Theoretical Physics, JINR, 141980, Dubna, Russia}
\author{Ch.~Stoyanov}
\email{stoyanov@inrne.bas.bg}
\affiliation{Institute for Nuclear Research and Nuclear Energy, Bulgarian Academy
of Sciences, 1784 Sofia, Bulgaria}

\date{\today}

\begin{abstract}
A microscopic thermodynamically consistent approach is applied to compute electron capture (EC) rates and cross sections on nuclei in hot stellar environments.
The cross section calculations are based on the Donnelly-Walecka multipole expansion method for treatment of semi-leptonic processes in nuclei.
To take into account thermal effects, we express the electron capture cross section in terms of temperature- and momentum-dependent spectral functions
for respective multipole charge-changing operators.  The spectral functions are computed by employing  the self-consistent thermal quasiparticle random-phase approximation (TQRPA) with the Skyrme effective interaction.
Three different Skyrme parametrizations (SkM$^*$, SGII and SLy4) are used to investigate thermal effects on EC for $^{56}$Fe and $^{78}$Ni. For $^{56}$Fe,  the impact of thermally unblocked
Gamow-Teller GT$_+$ transitions on EC is discussed and the results are compared with those from shell-model calculations. In particular,  it is shown that for some
temperature and density regimes the TQRPA rates exceed  the shell-model rates due to violation of the Brink-Axel hypothesis within the TQRPA. For neutron-rich $^{78}$Ni, the full momentum-dependence of
multipole transition operators is considered and it is  found that not only thermally unblocked allowed  $1^+$ transitions but also thermally unblocked first-forbidden $1^-$ and $2^-$ transitions  favor EC.
\end{abstract}



\maketitle

\section{Introduction}\label{intro}

The knowledge of low-energy nuclear weak-interaction-mediated processes is crucial for understanding the late stage of massive stars' evolution~\cite{Langanke_RevModPhys75,Janka_PhysRep442}.
Among them,  electron capture   strongly influences
the precollapse stage as well as the gravitational collapse of the iron core, leading to the supernova explosion. The collapse starts when the core
exceeds the  Chandrasekhar limit and electrons begin to be captured by iron-group nuclei. As electrons dominate the matter pressure,  the depletion of the electron population
due to capture by nuclei is a crucial factor determining the initial collapse phase. Until the core reaches  densities of $\rho\approx 10^{11}$ g\,cm$^{-3}$,
neutrinos produced by these reactions leave the star practically unhindered, cooling the core and reducing its entropy.
Moreover,  the  electron capture (EC) rates strongly determine the electron-to-baryon ratio $Y_e$  in a way that directly influences the collapse dynamics and the fate of the shock wave formed by the supernova explosion.
So, the nuclear electron capture is one of the most essential ingredients  involved in the complex dynamics of core-collapse supernova, and reliable
estimates of EC rates are crucial for better understanding of the explosion mechanism.

The determination of stellar EC rates  is a challenging nuclear structure
problem.   First of all, because of the low entropy in the core and the neutron-rich conditions, very neutron-rich nuclei may be produced with abundance several orders of
magnitude larger than that of free protons. Usually,  only theoretical weak interaction  rates for such nuclei are  available. Additionally,
in high-temperature stellar environments,   the  total EC rate is given by a sum of individual contributions $\lambda_i$ from
thermally excited states:
\begin{equation}
\lambda (T)= \sum_{i} p_i(T) \lambda_{i},
 \end{equation}
where $p_i(T)$ is the Boltzmann population factor for a parent state with energy~$E_i$ at temperature $T$.
The contributions from excited states remove the reaction threshold and at  high temperatures they dominate the EC rate.
However, the calculation of specific contributions $\lambda_{i}$  is a problem whose complexity grows considerably with temperature and for $T\approx 1$~MeV the state-by-state evolution
of individual contributions becomes computationally infeasible because of too many thermally populated states.

The first set of EC rates in stellar matter has been computed and published for $sd$- and $pf$-shell nuclei by Fuller  \textit{et al.}~\cite{Fuller_APJS42,Fuller_APJS48,Fuller_APJ252,Fuller_APJ293},
employing the independent particle model. The calculations were based on the idea by Bethe \textit{et al}.~\cite{Bethe_NPA324}, who first recognized the key role played by the Gamow-Teller (GT) resonance 
in stellar weak processes. With the improvement of nuclear structure models and computer algorithms, large-scale shell-model (LSSM) calculations have become possible for $pf$-shell  nuclei.
Their results on  GT strength distributions for iron-group nuclei  agreed quite well with experimental data~\cite{Caurier_NPA653}.
In~Refs.~\cite{Langanke_NPA673,Langanke_ADNDT79}, detailed shell-model calculations of the weak interaction rates for $pf$-shell nuclei up to $A=65$ were performed
and their incorporation into pre-supernova models~\cite{Heger_ApJ560,Heger_PRL86} demonstrated significant changes in the core entropy and the electron-to-baryon ratio $Y_e$.

Despite significant progress in computation capabilities, the straightforward extension of the LSSM approach to highly excited nuclear states and neutron-rich nuclei with $A>65$ still remains impossible 
due to the huge dimension of the model space involved. 
In~Refs.~\cite{Langanke_NPA673,Langanke_ADNDT79}, the first issue was overcome by employing the  Brink-Axel hypothesis, i.e., assuming that GT$_+$ strength distributions built on excited states are the 
same as for the nuclear ground state, but shifted by excitation 
energy.\footnote{The validity of the Brink-Axel hypothesis for the GT strength function is not obvious and its violation is confirmed by the shell-model Monte-Carlo studies at finite 
temperature~\cite{Radha_PRC56} and most recently by the shell-model calculations for $sd$-shell nuclei~\cite{Misch_PRC90}.}
To overcome the second problem, in Ref.~\cite{Langanke_PRC63} the so-called ``hybrid'' model was proposed. In this model, the rates are calculated using the random-phase approximation (RPA) built on an
average thermal nuclear state described by the Slater determinant
with temperature-dependent occupation numbers. The latter are  determined within the shell-model Monte Carlo (SMMC) approach, which accounts for both finite-temperature effects and correlations among
nucleons.  Using the hybrid model, Langanke \textit{et al}.~\cite{Langanke_PRL90} calculated electron capture rates for a sample of nuclei   with $A=66-112$ (the $pfg/sdg$ shell),
taking into account allowed (i.e., GT) and first-forbidden transitions.  In particular, it was found that  the electron capture on neutron-rich nuclei dominates over the capture on free protons, leading 
to significant changes in  the core collapse dynamics. Later, in Ref.~\cite{Juodagalvis_NPA848}, EC rates for more than 2200 neutron-rich nuclei  were produced using the same hybrid approach but 
utilizing the Fermi-Dirac parametrization for occupation factors.

The hybrid model clearly demonstrates the importance  of nuclear correlations that lead to configurational mixing and unblock  GT$_+$ transitions in neutron-rich nuclei. However,
because of the determinant form of the average thermal nuclear state, pairing correlations cannot be treated properly within the hybrid model. Furthermore, being based on the RPA, the hybrid model 
takes into account only an endoergic electron capture process and neglects de-excitation of thermally excited states of a parent nucleus.
To avoid these shortcomings and predict stellar weak-interaction rates for hot nuclei in a microscopic thermodynamically consistent way, the thermal quasiparticle random-phase approximation (TQRPA) was 
proposed in~Refs.~\cite{Dzhioev_PhAN72,Dzhioev_BRAS74,Dzhioev_PRC81}.  Unlike the approaches based on the shell-model, the TQRPA is formulated in  the grand-canonical ensemble and allows both energy 
and particle exchange between a nucleus and the stellar environment. Similar to the hybrid model,  the  TQRPA is based on a statistical formulation  of the nuclear many-body problem and enables 
one to obtain  a temperature-dependent strength function for $p\to n$ transitions involved in EC. However, in contrast to the hybrid model, the TQRPA makes it possible  to treat both endoergic and exoergic 
electron capture processes. Moreover, calculations performed in~Refs.~\cite{Dzhioev_PhAN72,Dzhioev_BRAS74,Dzhioev_PRC81} reveal the important thermal effects on GT$_+$ distributions in neutron-rich
 nuclei which occur due to destructive interference between thermal excitations and configurational mixing.  Namely, using the example of neutron-rich Ge isotopes,  it was shown that the weakening 
 of pairing correlations with temperature leads to a considerable ($\approx 8$\,MeV) downward shift of the GT$_+$ strength. As a result, the low-energy EC cross sections demonstrate a strong
  temperature dependence. No such effect was found in hybrid model calculations.

In~Refs.~\cite{Dzhioev_PhAN72,Dzhioev_BRAS74,Dzhioev_PRC81}, the TQRPA calculations for electron capture rates were performed with the phenomenological Hamiltonian of
the quasiparticle-phonon model (QPM)~\cite{Soloviev_1992}, whose parameters are adjusted locally, i.e., to properties of the nucleus under consideration.  
In~Refs.~\cite{Dzhioev_PhAN74,Dzhioev_PhAN77,Dzhioev_PRC89,Dzhioev_PRC92}, the same model Hamiltonian was used to study thermal effects on neutrino-nucleus reactions relevant to supernova 
simulations. To improve the predictive power of TQRPA calculations, in~Refs.\cite{Dzhioev_PRC94,Dzhioev_PhAN79} the method was combined with the Skyrme energy density functional theory. 
The resulting self-consistent Skyrme-TQRPA model can be used to make theoretical predictions for weak interaction processes with nuclei far from the stability valley more reliable. 
In the present work, we apply the Skyrme-TQRPA model to study stellar electron capture on nuclei in the iron-group mass region and for neutron-rich nuclei. To this aim, we perform EC calculations
for $^{56}$Fe and $^{78}$Ni. In~Refs.~\cite{Dzhioev_PhAN72,Dzhioev_BRAS74,Dzhioev_PRC81,Dzhioev_PhAN79}, the long wavelength approximation for allowed and first-forbidden transitions was used.  
This assumption is valid for low-energy electrons in the precollapse phase but it becomes doubtful at a later stage of the collapse when the increased density results in higher energy electrons 
($E_e \approx \rho^{1/3}$). To take into account   the full momentum dependence of transition operators, we employ the Donnelly-Walecka multipole expansion method  to treat semi-leptonic processes 
in nuclei~\cite{OConnell_PRC6,Walecka_1975} and express the EC cross section through  temperature- and momentum-dependent spectral functions.

We should mention several papers where different models based on RPA with the inclusion of temperature effects have been used to calculate stellar EC rates~\cite{Paar_PRC80,Fantina_PRC86, Niu_PRC83}. 
Our approach differs from those of Refs.~\cite{Paar_PRC80,Fantina_PRC86, Niu_PRC83} primarily by thermodynamically consistent consideration of thermal effects.
It was shown in Ref.~\cite{Dzhioev_PhAN79} that exoergic transitions from thermally excited states appear within the TQRPA and for EC on $^{56}$Fe
they remove the reaction threshold and enhance the low-energy cross section. In contrast, no such transitions appear within the finite-temperature RPA models.  As a result,
calculations in   Refs.~\cite{Paar_PRC80,Fantina_PRC86, Niu_PRC83} predict  that EC cross sections drop rapidly to zero as the electron energy falls below some threshold value.
We will return to this point in Sec.~\ref{results_for_56Fe}.

The paper is organized as follows: In Sec.~\ref{formalism}, the expressions necessary to calculate cross sections and rates of EC on hot nuclei are given. In addition, in Sec.~\ref{formalism}
we review the basics of the formalism and show how to compute charge-changing finite-temperature spectral functions within the TQRPA.
The results of the numerical calculations for $^{56}$Fe and $^{78}$Ni and their comparison with other models are presented and discussed in Sec.~\ref{results}. Conclusions are drawn in Sec.~\ref{conclusion}.

\section{Theoretical Formalism}
\label{formalism}

To compute EC rates in the hot supernova environment, we assume that the atoms are completely ionized and the
surrounding electron gas is described by the distribution function $f(E_e)$.
Then, neglecting the Pauli blocking for outgoing neutrinos, the stellar electron capture rate on a hot nucleus is obtained by folding the  finite-temperature cross 
section with the distribution of electrons
\begin{align}\label{rate_CrSec}
  \lambda(T) &=2\int\frac{d^3\pmb p_e}{(2\pi\hbar)^3} \sigma (E_e,T) c  f_e(E_e)
  \notag\\
  &= \frac{c}{\pi^2 (\hbar c)^3}\int\limits^\infty_{m_e c^2} \sigma (E_e,T) E_e p_e c f_e(E_e) d E_e,
\end{align}
where $p_e  = (E^2_e - m^2_ec^4)^{1/2}/c$ is the  momentum of the incoming electron with energy $E_e$.  Under conditions encountered in the collapsing core, the 
electron distribution is described by the Fermi-Dirac function with temperature $T$ and chemical potential $\mu_e$, i.e., $ f_e(E_e)\equiv f_e(E_e,\mu_e,T) $. 
The electron chemical potential $\mu_e$ is determined from the baryon density $\rho$ by inverting the relation
\begin{equation}
\rho Y_e = \frac{1}{\pi^2 N_A}\frac{1}{(\hbar c)^3}\int^\infty_0\bigl(f_{e}(E_e) - f_{p}(E_e)\bigr)(p_e c)^2d(p_e c),
\end{equation}
where $Y_e$ is the electron-to-baryon ratio and $N_A$ is the Avogadro constant. The positron
distribution function $f_p$ is defined by the substitution the chemical potential $\mu_p = -\mu_e$.

In~Eq.~\eqref{rate_CrSec}, the temperature-dependent cross section for  capture of an electron with energy $E_e$ is determined as
the following thermal average:
\begin{align}\label{sigmaET}
\sigma (E_e,T) &= \sum_{if} p_i(T) \int d\mathrm{\Omega} \frac{d\sigma_{i\to f}(E_e)}{d\mathrm{\Omega}}
\notag\\
&=\int\limits^{E_e}_{-\infty}dE\int d \mathrm{\Omega} \frac{d^2\sigma(E_e,E, T)}{dEd\mathrm{\Omega}},
\end{align}
while the finite-temperature differential cross section is defined as
\begin{equation}\label{DDCS1}
  \frac{d^2\sigma(E_e,E, T)}{dEd\mathrm{\Omega}} = \sum_{if}p_i(T)\frac{d\sigma_{i\to f}(E_e)}{d\mathrm{\Omega}}\delta(E-\Delta E_{fi}).
\end{equation}
Here,  $E =  E_e-E_\nu$ is the energy transferred to the nucleus when emitting the neutrino with energy $E_\nu$.
In the above definitions we account for all energetically allowed transitions, i.e., $\Delta E_{fi} \le E_e$, where
$\Delta E_{fi}$ is the transition energy needed to go from the  parent nuclear state $i$ to the daughter nuclear state $f$. For proton-to-neutron transitions
$\Delta E_{fi}=\varepsilon_f-\varepsilon_i +\Delta M_{np}$, where $\Delta M_{np}=1.293$~MeV is the neutron-proton mass difference and
$\varepsilon_{i(f)}=\langle i(f)|H| i(f) \rangle$ with  $H$ being the nuclear Hamiltonian. The important point is that due  to thermally excited states the energy transfer $E$  can 
 be both positive and negative.

In the derivation of the temperature-dependent EC cross section, we  follow the Donelly-Walecka formalism~\cite{OConnell_PRC6,Walecka_1975} (see also~Ref.~\cite{Niu_PRC83}), which is based on the standard
current-current form of the weak interaction Hamiltonian. Applying multipole expansion of the weak hadronic current, the method allows one to express
the electron-nucleus differential cross section in Eq.~\eqref{DDCS1}  through the matrix elements of the charge $\hat M_J$, longitudinal $\hat L_J$,  transverse
electric $\hat T^\mathrm{el}_J$, and transverse magnetic $\hat T^\mathrm{mag}_J$ operators. Then, the  differential cross section~\eqref{DDCS1} can be written as the following multipole expansion:
  \begin{align}\label{DDCS2}
  \frac{d^2\sigma(E_e,E, T)}{dEd\mathrm{\Omega}}&= \frac{(G_\mathrm{F} V_{ud})^2}{2\pi(\hbar c)^4}\,E^2_\nu\,{\mathcal{R}(E_e, E_\nu)} F(Z, E_e)
  \notag\\
   &\times\Bigl\{\sum_{J=0}^{\infty}\sigma^J_{CL}(E,T) + \sum_{J=1}^\infty\sigma^J_{T}(E,T)\Bigr\}.
  \end{align}
Here, $G_\mathrm{F}$ is the Fermi coupling constant and  $V_{ud}$ is the up-down element in the Cabibbo-Kobayashi-Maskawa  quark mixing matrix.   
The Fermi function $F(Z,E_e)$ corrects the cross section for the distortion of the electron wave function by the Coulomb field of the nucleus~\cite{Langanke_NPA673},
 while the factor ${\mathcal{R}(E_e, E_\nu)}$ accounts for
the nuclear recoil~\cite{Niu_PRC83}.\footnote{For  relevant electron energies,  ${\mathcal{R}(E_e, E_\nu)}\approx 1$.}

In~Eq.~\eqref{DDCS2}, all temperature dependence is contained in the Coulomb, longitudinal and transverse  multipole components
\begin{align}\label{CLT}
  \sigma^J_{CL, T}(E,T)=\sum_{if}p_i(T)\sigma^{J}_{CL,T}(i\to f)\delta(E-\Delta E_{fi}).
\end{align}
For spherical nuclei, the explicit expressions for $\sigma^{J}_{CL,T}(i\to f)$ through the reduced matrix elements of
all the above multipole operators
are given in~Refs.~\cite{OConnell_PRC6,Walecka_1975}. These matrix elements depend on the four-momentum transfer $(E,\pmb q)$ to the nucleus and they also include the nucleon vector, axial-vector, and
pseudoscalar form factors~\cite{Paar_PRC80}. Substituting the expressions for $\sigma^{J}_{CL,T}(i\to f)$ into Eq.~\eqref{CLT}, we express the temperature-dependent
components  $\sigma^J_{CL, T}(E,T)$ through the  spectral  functions for charge, longitudinal,  transverse electric, and transverse magnetic multipole operators:
\begin{multline}\label{sigmaCL}
  \sigma^J_{CL}(E,T) = (1+a\cos\Theta)S_{M_JM_J}
  \\
  + (1+a\cos\Theta - 2b\sin^2\Theta)S_{L_JL_J}
  \\
  +\Bigl[\frac{E}{q}(1+a\cos\Theta)+c\Bigr]2\mathrm{Re}\{S_{M_JL_J} \}
\end{multline}
and
\begin{multline}\label{sigmaT}
  \sigma^J_{T}(E,T) = (1-a\cos\Theta+b\sin^2\Theta)\bigl[S_{T^\mathrm{mag}_J T^\mathrm{mag}_J} + S_{T^\mathrm{el}_J T^\mathrm{el}_J} \bigr]
  \\
   -\Bigl[\frac{E_e+E_\nu}{q}(1-a\cos\Theta)-c\Bigr]2\mathrm{Re}\{S_{T^\mathrm{mag}_J T^\mathrm{el}_J} \}.
\end{multline}
The following notation is used above:
\begin{equation}
  a = \sqrt{1-\Bigl(\frac{m_ec^2}{E_e}\Bigr)},~~b = \frac{E_e E_\nu a^2}{q^2},~~c=\frac{(m_ec^2)^2}{qE_e},
\end{equation}
and the absolute value of the three-momentum transfer depends on the scattering angle $\Theta$ as
\begin{equation}
  q = |\vec q| = \sqrt{E^2 + 2E_e E_\nu(1-a\cos\Theta)-(m_ec^2)^2}.
\end{equation}
In Eqs.~\eqref{sigmaCL} and \eqref{sigmaT}, the spectral  function $ S_{A_{J}B_{J}}(E,T)$ for   multipole operators $A_{JM}$ and $B_{JM}$ is defined as
\begin{multline}\label{p-to-n}
  S_{A_{J}B_{J}}(E,T) =
  \\
  \sum_{if} p_i(T)\frac{\langle J_f \| B_J\|J_i\rangle\langle J_f\| A_J\| J_i\rangle^*}{2J_i+1} \delta(E-\Delta E_{fi}),
\end{multline}
where $J_{i(f)}$ is the angular momentum of the initial (final) nuclear state. Because of transitions from thermally excited states, the spectral functions are determined for both
positive and negative energies.

For low-energy electrons, when the long wavelength limit $q\to 0$ is valid,  the structures of the $1^+$ multipole operators entering into $\sigma^J_{CL}$ and $\sigma^J_T$ ($J^\pi = 1^+$) 
reduce to the Gamow-Teller form $\mathrm{GT}_+= g_A\pmb\sigma t_+$~\cite{Donnelly_PhR50}, where $g_A$ is the axial-vector coupling constant. Then,
the $1^+$ component of the cross section takes the form
\begin{align}\label{CrSect_GT}
  \sigma_\mathrm{GT}(E_e,  T) =&  \frac{ (G_\mathrm{F} g_V V_\mathrm{ud})^2}{2\pi(\hbar c)^4} F(Z,E_e)
  \notag \\
  &\times\int^{E_e}_\infty(E-E_e)^2 S_\mathrm{GT}(E,T) dE,
\end{align}
where $g_V$ is  the vector coupling constant, whereas  the temperature dependent strength function for the GT$_+$ transition operator is defined as
\begin{equation}\label{SGT1}
   S_\mathrm{GT}(E,T) = \Bigl(\frac{g_A}{g_V}\Bigr)^2\sum_{if} p_i(T)\frac{\bigl|\langle J_f\|\pmb\sigma t_+\| J_i\rangle\bigr|^2}{2J_i+1}\delta(E-\Delta E_{fi}),
\end{equation}
where $g_A/g_V = -1.27$. By substituting~\eqref{CrSect_GT}
into~\eqref{rate_CrSec} we get the  Gamow-Teller contribution $\lambda_\mathrm{GT}$ to the EC rate.

To compute the spectral  functions, we consider the nuclei embedded in a hot and dense presupernova medium as
open quantum systems in thermal equilibrium with heat and particle
reservoirs and, hence, they can be described as a thermal grand-canonical ensemble with temperature $T$ and
chemical potentials of protons $\lambda_p$ and neutrons $\lambda_n$, respectively. The grand-canonical probability distribution $p_i(T)\equiv P(\varepsilon_i, A^Z_N)$ is given by
\begin{equation}
  P(\varepsilon_i, A^Z_N)=(2J_i+1)\exp\Bigl\{-\frac{\varepsilon_i -\lambda_n N-\lambda_p Z}{T}\Bigr\}\Bigl/\mathcal{Z}(T),
\end{equation}
where $\mathcal{Z}$ is the  partition function. Within the grand-canonical ensemble,
the spectral  function for charge-changing $p\to n$ transition  operators can be written as the Fourier transform of the
time-correlation function
\begin{multline}\label{SCF1}
  S_{A_{J}B_{J}}(E,T) = \int \frac{dt}{2\pi}\,\mathrm{e}^{i(E-\delta_{np})t}\sum_M\langle\!\langle A^\dag_{JM}(t) B_{JM}(0)\rangle\!\rangle,
\end{multline}
where $\delta_{np} =\Delta M_{np}+\Delta\lambda_{np}$ with $\Delta\lambda_{np}=\lambda_n-\lambda_p$, and $ A_{JM}(t) = \mathrm{e}^{iH't}A_{JM}\mathrm{e}^{-iH't} $
 with $H'=H-\lambda_n\hat N-\lambda_p\hat Z$. The double brackets $\langle\!\langle\ldots\rangle\!\rangle$ mean
the grand-canonical average, i.e.,
\begin{equation}
 \langle\!\langle \mathcal{O} \rangle\!\rangle \equiv \sum_{N,Z}\sum_{i, M_i}(2J_i+1)^{-1} P(\varepsilon_i, A^Z_N)\langle J_i M_i| \mathcal{O}| J_i M_i\rangle.
\end{equation}
The grand-canonical time-correlation function in~\eqref{SCF1} satisfies the Kubo-Martin-Schwinger (KMS) condition~\cite{Kubo_JPSJ12,Martin_PRev115}
\begin{equation}\label{KMS}
\langle\!\langle A^\dag(t) B(0)\rangle\!\rangle = \langle\!\langle B(0)A^\dag(t+i\beta)\rangle\!\rangle,~~~(\beta=1/T).
\end{equation}
Then, elementary calculations show that the spectral function~\eqref{SCF1} is connected to
 the spectral function for Hermitian conjugate $n\to p $ operators $A^\dag_{JM},~B^\dag_{JM}$ by the following detailed balance relation:
 \begin{equation}\label{DBR}
 S_{B^\dag_{J}A^\dag_{J}}(-E,T)  = e^{-(E-\delta_{np})/T}S_{A_{J}B_{J}}(E,T),
\end{equation}
where
\begin{multline}\label{SCF2}
  S_{A^\dag_{J}B^\dag_{J}}(E,T) = \int \frac{dt}{2\pi}\,e^{i(E+\delta_{np})t}\sum_M\langle\!\langle A_{JM}(t) B^\dag_{JM}(0)\rangle\!\rangle.
\end{multline}
Note a different sign before $\delta_{np}$ in comparison with Eq.~\eqref{SCF1}. It must be emphasized  that in the form~\eqref{DBR},  the detailed balance  
for charge-changing spectral functions is valid only within the grand-canonical ensemble. Within the canonical ensemble, the detailed balance for charge-changing processes 
was derived in Ref.~\cite{Misch_APJ844} and it involves partition functions for the parent and daughter nuclei.

So the problem of computing the electron capture cross sections and rates on hot nuclei is reduced to determining the time-correlation functions
for charge-changing multipole operators $\hat M_J$, $\hat L_J$,  $\hat T^\mathrm{el}_J$, and $\hat T^\mathrm{mag}_J$.
To compute $\langle\!\langle A(t)B(0)\rangle\!\rangle$,  we apply  the formalism, which is called the thermo field dynamics (TFD).
The concept of TFD is expounded  in~Refs.~\cite{Takahashi_IJMPB10,Umezawa1982,Ojima_AnPhys137}, and here we only outline the key points relevant to the present discussion.

Formally, the TFD approach stems from the possibility of writing the statistical average $\langle\!\langle \mathcal{O} \rangle\!\rangle$  as an expectation value over a temperature-dependent 
state $|0(T)\rangle$ called the thermal vacuum,
\begin{equation}\label{average}
 \langle\!\langle \mathcal{O} \rangle\!\rangle \equiv \langle 0(T)|\mathcal{O} |0(T)\rangle.
\end{equation}
In this sense, the thermal vacuum describes the  system in the thermal equilibrium.
In order to define  $|0(T)\rangle$, one needs
to double  the original Hilbert space by introducing a fictitious dynamical system, identical to the initial one. The doubling of the Hilbert space, which
is the doubling of the states, then involves doubling the Hamiltonian
of the system.  Let $H=H(a^\dag,a)$ be the nuclear Hamiltonian.  If we denote the fictitious quantities by the tilde, then
the Hamiltonian of the  fictitious dynamical system has the form $\widetilde H=H(\widetilde a^\dag,\widetilde a)$.
Physically, the origin of tilde creation and annihilation operators can be seen as the result of the interaction
between the system with the surrounding thermal reservoir, the latter maintaining a certain number of excited quanta in 
the system.\footnote{The correspondence between the thermo field dynamics and the superoperator formalism is discussed in~Ref.\cite{Schmutz_ZPhysB30}.}
Then, doubling of the system degrees of freedom allows us to consider excitation and de-excitation processes at finite temperature.
To ensure~\eqref{average}, the thermal vacuum should satisfy two properties: (i)  $|0(T)\rangle$  is
 the zero-energy eigenstate of the so-called thermal Hamiltonian $\mathcal{H} = H - \widetilde H$, i.e.,
$\mathcal{H}  |0(T)\rangle = 0$; and (ii) the following thermal state-condition is valid for  an arbitrary operator $A$:
\begin{equation}\label{TSC}
  A|0(T)\rangle = \sigma_A \mathrm{e}^{\mathcal{H}/2T}\widetilde A^\dag |0(T)\rangle,
\end{equation}
where $\sigma_A$ is a phase factor and the correspondence between $A$ and $\widetilde A$ is given by the tilde-conjugation rules~\cite{Takahashi_IJMPB10,Umezawa1982,Ojima_AnPhys137}. 
It is shown in~Ref.~\cite{Umezawa1982} that Eq.~\eqref{TSC} is  equivalent to the KMS condition~\eqref{KMS}.

To demonstrate how to compute the spectral functions within the TFD, we first replace the time-correlation function in Eq.~\eqref{SCF1} by the
thermal vacuum expectation value
\begin{align}\label{SCF2}
  S_{A_{J}B_{J}}(E,T) &= \int \frac{dt}{2\pi}\, \mathrm{e}^{i(E-\delta_{np})t}
  \notag\\
  &\times\sum_M \langle 0(T)| A^\dag_{JM}(t)B_{JM}(0)|0(T)\rangle.
\end{align}
Since $\widetilde H$ contains an even number of tilde creation and annihilation operators, it  commutes with all physical operators. Therefore, we can formally
write
\begin{equation}
  A(t) = \mathrm{e}^{i\mathcal{H}t} A\, \mathrm{e}^{-i\mathcal{H}t}.
\end{equation}
Let us now assume that we can find eigenstates and eigenvalues of the thermal Hamiltonian
\begin{align}\label{exactESEV}
  \mathcal H|\mathrm{\Psi}_k\rangle = \varepsilon_k(T)  |\mathrm{\Psi}_k\rangle,~~~
   \mathcal H|\mathrm{\widetilde\Psi}_k\rangle = -\varepsilon_k(T)  |\mathrm{\widetilde\Psi}_k\rangle
\end{align}
and $\langle 0(T)|\mathrm{\Psi}_k\rangle=\langle 0(T)|\mathrm{\widetilde\Psi}_k\rangle=0$. Note that temperature-dependent eigenstates of $\mathcal H$ form pairs: For each  $|\mathrm{\Psi}_k\rangle$
with the eigenvalue $\varepsilon_k(T)>0$,   there is a tilde-conjugated state  $|\mathrm{\widetilde\Psi}_k\rangle$ which is also an eigenstate of $\mathcal H$ with the
eigenvalue $-\varepsilon_k(T)$.
Because of the completeness of the thermal  Hamiltonian eigenstates, we can rewrite  Eq.~\eqref{SCF2} in the following form:
  \begin{align}\label{str_funct2}
   & S_{A_JB_J}(E,T) = \sum_{M,k}\Bigl\{ \bigl\langle\mathrm{\Psi}_k | B_{JM} |0(T)\bigr\rangle  \bigl\langle \mathrm{\Psi}_k | A_{JM} |0(T)\bigr\rangle^*
    \notag\\
   & \times  \delta(E-\delta_{np}-\varepsilon_{k})+\bigl\langle\mathrm{\widetilde\Psi}_k | B_{JM} |0(T)\bigr\rangle \bigl\langle \mathrm{\widetilde\Psi}_k | A_{JM} |0(T)\bigr\rangle^*
   \notag \\
   &\times \delta(E-\delta_{np}+\varepsilon_{k})\Bigr\}.
  \end{align}
 Thus, within the TFD the spectral function is expressed through the transition matrix elements of the operators $A_{JM}$ and $B_{JM}$
 taken between the thermal vacuum and eigenstates of the thermal Hamiltonian $\mathcal{H}$.
 The singularities of the spectral function  correspond to temperature-dependent  eigenvalues of the thermal Hamiltonian shifted by the value of $\delta_{np}$.
 At $T=0$, the transition matrix elements to tilde states are zero and, therefore,  $S_{A_JB_J}(E,T=0)$  is nonvanishing only for $E>\delta_{np}$. So we can think
 about $\delta_{np}$ as an ``effective'' ground-states threshold for $p\to n$ reactions. For  $n\to p$ reactions, the ``effective'' ground-state threshold is $-\delta_{np}$.
  At finite temperature, $S_{A_JB_J}(E,T)$ is nonzero for both  $E>\delta_{np}$ and  $E<\delta_{np}$  energies and the   latter describe  de-excitation processes of a hot system, i.e., 
transition from high-energy thermally excited states to states at lower energies. Using the thermal state condition~\eqref{TSC} and
 taking into account the property $\langle \mathrm{\Psi}_k|\widetilde A|0(T)\rangle^* = \langle \widetilde{\mathrm\Psi}_k | A|0(T)\rangle$, we easily derive the detailed balance relation~\eqref{DBR}  
from Eq.~\eqref{str_funct2}.

From the above considerations, it becomes clear how to use the TFD to compute the EC  rates and cross sections for hot nuclei: This is
the diagonalization of the thermal nuclear Hamiltonian and the subsequent computation of spectral functions. Obviously, in most practical cases  we cannot diagonalize $\mathcal{H}$
exactly and find the exact thermal vacuum state. The merit of TFD, however, allows one to resort to approximations valid at zero temperature. Hence, the
thermal vacuum can be constructed in the Hartree-Fock-Bogoliubov approximation, or in the random phase approximation. Moreover, the concept of quasiparticles and phonons can
be extended to $T\ne 0 $ within the TFD and the thermal vacuum can be defined as the vacuum state for respective annihilation operators~\cite{Dzhioev_IJMPhE18}.

In the present work, we compute the spectral functions by applying  the so-called thermal quasiparticle RPA method. For charge-changing transitions in hot nuclei,
the TQRPA was introduced in Refs.~\cite{Dzhioev_PhAN72,Dzhioev_BRAS74,Dzhioev_PRC81, Dzhioev_PhAN79}. Let us, for the sake of completeness, briefly recall the method.
Within the TQRPA,  eigenstates of the thermal Hamiltonian are treated as phonon-like excitations on the thermal vacuum
\begin{align}\label{ph_ex}
  &|Q_{JMi}\rangle=Q^\dag_{JMi}|0(T)\rangle,
  \notag\\
  &|\widetilde Q_{JMi}\rangle=\widetilde Q^\dag_{\overline{JM}i}|0(T)\rangle,
\end{align}
where we denote $\widetilde Q^\dag_{\overline{JM}i}=(-1)^{J-M}\widetilde Q^\dag_{J-Mi}$, while nontilde and tilde-phonon operators are connected by the
tilde-conjugation rules~\cite{Takahashi_IJMPB10,Umezawa1982,Ojima_AnPhys137}. The thermal vacuum itself is a vacuum for the $Q_{JMi}$ and $\widetilde Q_{JMi}$ operators. 
We apply the TQRPA to the general nuclear Hamiltonian containing a spherical mean field for protons and neutrons, pairing and residual particle-hole interactions
\begin{equation}\label{nucl_Hamiltonian}
  H = H_\mathrm{mf}+H_\mathrm{pair} + H_\mathrm{ph}.
\end{equation}
Since we are working in the the grand-canonical ensemble, the chemical potentials for protons and neutrons $\lambda_p$ and $\lambda_n$ are included into $H_\mathrm{mf}$.  Following the TFD prescription, we construct the thermal Hamiltonian and then approximately diagonalize it using the same techniques and approximations as for a ``cold'' nucleus.
Namely, we first  introduce thermal quasiparticles that diagonalize the mean field and pairing parts of $\mathcal{H}$:
\begin{equation}\label{H_pairing}
  \mathcal{H}_\mathrm{mf}+\mathcal{H}_\mathrm{pair}\simeq\sum_\tau{\sum_{jm}}^\tau\varepsilon_{j}(T)
  (\beta^\dag_{jm}\beta_{jm} -
  \widetilde\beta^\dag_{j m}\widetilde\beta_{jm})
\end{equation}
and their vacuum is the thermal vacuum in the BCS approximation. In the expression above, the notation ${\sum}^\tau$ implies a summation over neutron 
($\tau=n$) or proton ($\tau=p)$ single particle states only.   The energy and the structure of thermal quasiparticles
are found from the finite-temperature BCS equations. At the next step, we take into account the residual particle-hole interaction and
diagonalize   $\mathcal{H}$ in terms of thermal multipole phonons,
\begin{equation}\label{thermal_H_phonon}
  \mathcal{H}\simeq \sum_{JMi}\omega_{Ji}(T)(Q^\dag_{JMi}Q_{JMi}-\widetilde Q^\dag_{JMi}\widetilde Q_{JMi}).
\end{equation}
For charge-changing processes like electron capture or $\beta$-decay, the thermal phonon operators are constructed as a linear
superposition of the creation and annihilation operators for proton-neutron thermal quasiparticle pairs,
\begin{multline}\label{phonon}
  Q^\dag_{J M i}=\sum_{j_pj_n}
 \Bigl\{\psi^{Ji}_{j_pj_n}[\beta^\dag_{j_p}\beta^\dag_{j_n}]^J_M +
 \widetilde\psi^{J i}_{j_pj_p}[\widetilde\beta^\dag_{\overline{\jmath_p}}
 \widetilde\beta^\dag_{\overline{\jmath_n}}]^J_M
 \\ +
 i\eta^{J i}_{j_pj_n}[\beta^\dag_{j_p}
  \widetilde\beta^\dag_{\overline{\jmath_n}}]^J_M
+
 i\widetilde \eta^{J i}_{j_pj_n}[\widetilde\beta^\dag_{\overline{\jmath_p}}
\beta^\dag_{j_n}]^J_M
 \\ +
 \phi^{J i}_{j_pj_n}[\beta_{\overline{\jmath_p}}\beta_{\overline{\jmath_n}}]^J_M
+ \widetilde\phi^{J i}_{j_pj_n}[\widetilde\beta_{j_p}
 \widetilde\beta_{j_n}]^J_M \\+
  i\xi^{J i}_{j_pj_n}[\beta_{\overline{\jmath_p}}
  \widetilde\beta_{j_n}]^J_M
 +
  i\widetilde\xi^{J i}_{j_pj_n}[\widetilde\beta_{j_p}
\beta_{\overline{\jmath_n}}]^J_M
  \Bigr\},
\end{multline}
with $[\,\,]^J_M$ denoting the coupling of two angular momenta $j_p$ and $j_n$ to the total angular momentum $J$ and its projection $M$.
As an additional constraint, we demand  that the vacuum of thermal phonons obeys the thermal-state condition~\eqref{TSC}. Then,  the energy and the structure of thermal charge-changing phonons 
are obtained by the solution of the TQRPA equations.  In the zero-temperature limit, the TQRPA method reduces to the standard QRPA.

To clarify the physical meaning of different terms in~\eqref{phonon}, we note that the creation of a negative-energy tilde thermal quasiparticle corresponds to the annihilation 
of a thermally excited Bogoliubov quasiparticle or, which is the same, to the creation of a quasihole state (see~Ref.~\cite{Dzhioev_PRC81} for more details). Therefore, at finite temperature, 
charge-changing single-particle transitions involve excitations of three types: (1) two-quasiparticle excitations described by the operator $\beta^\dag_{j_p}\beta^\dag_{j_n}$ and having energy 
$\varepsilon^{(+)}_{j_pj_n}=\varepsilon_{j_p}+\varepsilon_{j_n}$, (2) one-quasiparticle--one-quasihole excitations
described by the operators $\beta^\dag_{j_p}\widetilde\beta^\dag_{j_n}$, $\widetilde\beta^\dag_{j_p}\beta^\dag_{j_n}$ and having energies $\varepsilon^{(-)}_{j_pj_n}=\varepsilon_{j_p}-\varepsilon_{j_n}$
 and $-\varepsilon^{(-)}_{j_pj_n}$, respectively, and (3) two-quasihole excitations described by the operator $\widetilde\beta^\dag_{j_p}\widetilde\beta^\dag_{j_n}$ and having energy
$-\varepsilon^{(+)}_{j_pj_n}$. The last two types are possible only at $T\ne0$. Therefore, due to single-particle
transitions involving annihilation of thermally excited Bogoliubov quasiparticles, the phonon spectrum at finite temperature contains negative- and low-energy states which do not exist at zero 
temperature and these ``new'' phonon states can be interpreted as thermally unblocked transitions between nuclear excited states. In~\eqref{thermal_H_phonon}, thermal phonons with negative energy 
are denoted by tilde and they  contribute to the spectral function at $E<\delta_{np}$.

Once the energy and the structure of thermal phonons are determined, the TQRPA spectral function for the $p\to n$ multipole
operators  can be written as
   \begin{multline}\label{str_funct3}
    S_{A_JB_J}(E,T) =
    \\
     \sum_{i} \bigl\langle Q_{Ji}\| B_J \|0(T)\bigr\rangle \bigl\langle Q_{Ji}\| A_J \|0(T)\bigr\rangle^* \delta(E-E_{Ji}),
    \\
    +\sum_{i} \bigl\langle \widetilde Q_{Ji}\| B_J \|0(T)\bigr\rangle
     \bigl\langle\widetilde Q_{Ji}\| A_J \|0(T)\bigr\rangle^* \delta(E-\widetilde E_{Ji}),
 \end{multline}
where the following notation for ``reduced'' transition matrix elements is used:
 \begin{equation}
   \bigl\langle Q_{Ji}\| B_J \|0(T)\bigr\rangle \equiv (2J+1)^{1/2}\bigl\langle Q_{JMi}| B_{JM} |0(T)\bigr\rangle.
 \end{equation}
The transition energies to nontilde and tilde phonon states are given by
\begin{equation}\label{trEnergies}
  E_{Ji} = \omega_{Ji}(T)+\delta_{np},~~\widetilde E_{Ji} = -\omega_{Ji}(T)+\delta_{np},
\end{equation}
while detailed expressions for the transition matrix elements of the charge-changing multipole operators are given in Ref.~\cite{Dzhioev_PhAN72}. 
The resulting spectral function is temperature-dependent since both the matrix elements and transition energies depend on temperature. An additional point
 to emphasize is that the spectral functions computed within the TQRPA obey the detailed balance relation~\eqref{DBR}.
In this sense, the present version of the TQRPA is a thermodynamically  consistent framework.

Given the spectral functions for the multipole operators $\hat M_J$, $\hat L_J$,  $\hat T^\mathrm{el}_J$, and $\hat T^\mathrm{mag}_J$ and substituting the results into 
Eqs.~\eqref{sigmaCL} and~\eqref{sigmaT},  we  get the  Coulomb (longitudinal) and transverse multipole components $\sigma^J_{CL}$,  $\sigma^J_{T}$ which determine the EC cross  sections and rates.
Restricting our consideration to low-energy $1^+$ transitions, i.e., neglecting momentum transfer, we can derive explicit expressions for $\lambda_\mathrm{GT}$ and $\sigma_\mathrm{GT}$. Namely,
for the GT$_+$ strength function, only transitions from the thermal vacuum to thermal phonon states with $J^\pi=1^+$ are relevant. Then, omitting the index $J$,
the strength function $S_\mathrm{GT}$~\eqref{SGT1} can be written as
 \begin{equation}\label{strengthGT}
   S_\mathrm{GT}(E,T) = \sum_i \bigl\{ B^{(+)}_i \delta(E-E_{i}) + \widetilde B^{(+)}_i \delta(E-\widetilde E_{i})\bigr\},
 \end{equation}
where the GT$_+$ matrix element $B^{(+)}_i$ is given by
\begin{equation}\label{BGT_pl}
  B^{(+)}_i = \biggl(\frac{g_A}{g_V}\biggr)^2\bigl|\langle Q_{i}\| \pmb\sigma t_+\|0(T)\rangle\bigr|^2,
\end{equation}
while $\widetilde B^{(+)}_i$ is obtained from $B^{(+)}_i$ by replacing the nontilde state by  a tilde one. From~\eqref{strengthGT} it follows  that the GT$_+$ strength
below (above) $\delta_{np}$ corresponds to transitions to (non)tilde one-phonon states.
By substituting Eq.~\eqref{strengthGT} into Eq.~\eqref{sigmaET}, we get  the EC cross section due to GT$_+$ transitions:
\begin{multline}\label{CrSect_GT2}
  \sigma_\mathrm{GT}(E_e,  T) =  \frac{ (G_\mathrm{F} g_V V_\mathrm{ud})^2}{2\pi(\hbar c)^2} F(Z,E_e)
  \\
  \times\sum_i\Bigl\{(E_e-E_{i})^2 B_i + (E_e-\widetilde E_{i})^2\widetilde B_i\Bigr\},
\end{multline}
where summation is performed over $J^\pi=1^+$ thermal one-phonon states with transition energy $E_{i},~\widetilde E_{i} \le E_e$. Then, for the EC rate we get
\begin{equation}\label{rate_GT}
  \lambda_\mathrm{GT}=\frac{\ln2}{K}\sum_i\bigl\{ B_i \Phi^\mathrm{ec}(E_i) + \widetilde B_i \Phi^\mathrm{ec}(\widetilde E_i)\bigr\},
\end{equation}
where
\begin{equation}
  K=\frac{2\pi^3(\ln2)\hbar^7}{G^2_\mathrm{F}V^2_\mathrm{ud} g^2_V m^5_e c^4}=6150~\mathrm{s},
  \end{equation}
while the  phase space integral  $\Phi^\mathrm{ec}(E)$ is   given by
\begin{equation}
\Phi^\mathrm{ec}(E) = \frac{1}{(m_e c^2)^5}\int\limits^\infty_{E_\mathrm{min}}F(Z,E_e)(E_e-E)^2 E_e p_e c f_e(E_e)dE_e.
\end{equation}
Here, the capture threshold is $E_\mathrm{min}=m_ec^2$ if $E<0$ and $E_\mathrm{min}=E$ otherwise.

To finish this part, we would like to mention some important properties concerning the GT matrix elements.  In the same manner as above [see Eq.~\eqref{BGT_pl}] we define the reduced matrix elements
for the GT$_-$ operator
\begin{equation}
  B^{(-)}_i = \biggl(\frac{g_A}{g_V}\biggr)^2\bigl|\langle Q_{i}\| \pmb\sigma t_-\|0(T)\rangle\bigr|^2.
\end{equation}
Using the value $\widetilde B^{(-)}_i$ which refers to GT$_-$ transitions to  tilde-phonon states, we can write
the detailed balance relation~\eqref{DBR} connecting the GT$_+$ and GT$_-$ matrix elements as
\begin{equation}\label{DBR2}
   \widetilde B^{(\mp)}_i=\mathrm{e}^{-\omega_{i}(T)/T} B^{(\pm)}_i.
\end{equation}
Thus, for each $p\to n$ ($n\to p$) transition to a non-tilde thermal phonon state with $\omega_i>0$, there is an inverse $n\to p$ ($p\to n$) transition to a
tilde-conjugated state with energy $-\omega_i$.

One more important point is that the TQRPA approach preserves the Ikeda sum rule for the bare operators $\mathrm{GT}_\pm=\pmb\sigma t_\pm$
\begin{equation}\label{Ikeda}
  \sum_i\bigl\{ B^{(-)}_i + \widetilde B^{(-)}_i\bigr\} -   \sum_i\bigl\{ B^{(+)}_i + \widetilde B^{(+)}_i\bigr\} = 3(N-Z),
\end{equation}
although individually the total GT$_{\pm}$ strengths may vary with temperature. This was shown in Ref.~\cite{Dzhioev_PhAN72}.

\section{Results}\label{results}

In this section, we apply  the theoretical framework described above to evaluate EC rates on the two sample nuclei, $^{56}\mathrm{Fe}$ and $^{78}\mathrm{Ni}$. The iron isotope
is among the nuclei with $A\approx 60$, which are mainly responsible for decreasing the electron-to-baryon ratio during the oxygen- and silicon-burning phases of the massive star evolution. 
Therefore, the calculated EC rates for nuclei in this mass range are essential for modeling the initial phase of the stellar core collapse~\cite{Langanke_RevModPhys75,Janka_PhysRep442,Bethe_NPA324}.
Very neutron-rich nuclei dominate the nuclear composition during the last phase of the collapse before neutrino trapping.  It was shown in a recent work by 
Sullivan~\textit{et al}.~\cite{Sullivan_APJ816} that nuclei around the $N=50$ closed neutron shell at and above $^{78}$Ni have the largest impact on the change in the electron density $Y_e$  and, thus,
on the dynamical evolution of the collapse.

To compute the spectral functions within the TQRPA, we suppose that the Hartree-Fock mean-field potential $H_\mathrm{mf}$ and the residual particle-hole
interaction $H_\mathrm{ph}$ in Eq.~\eqref{nucl_Hamiltonian} are obtained self-consistently from the same Skyrme energy density functional. Moreover, we neglect the thermal effects on the Hartree-Fock 
single-particle states.  According to~Ref.~\cite{Bortignon_1998}, the stability of the mean field $H_\mathrm{mf}$ with respect to temperature is expected
for $T$ values considerably smaller than the energy difference between the
major shells ($\hbar\omega_0 = 41A^{-1/3})$. This requirement is well satisfied in nuclei with $A < 100$ for maximum temperatures
reached during the collapse ($T\approx 5$~MeV).  The single-particle continuum is discretized  by diagonalizing the HF Hamiltonian on a basis of 12 harmonic oscillator shells and cutting off
the single-particle spectra at the energy of 50~MeV.

The  isovector part of the residual particle-hole interaction $H_\mathrm{ph}$ is defined in terms of second derivatives of the Skyrme energy density
functional with respect to the one-body density~\cite{Bertsch_PhRep18}. In the present study, we neglect the spin-orbit part of the residual interaction.
Consequently, our calculations are not, strictly speaking, fully self-consistent.
However, as pointed out in Ref.~\cite{Fracasso_PRC76}, the  spin-orbit residual interaction practically has no effect on the GT excitations.
The same conclusion was made more recently in~Rev.~\cite{Wen2013}.
Following~Ref.~\cite{Giai_PRC57}, we then simplify the full velocity-dependent $H_\mathrm{ph}$  by approximating it by its
Landau–Migdal form in the momentum space,
\begin{equation}\label{S_Landau_momentum}
  H_\mathrm{ph}(\pmb k_1,\pmb k_2)=N_0^{-1}\sum_l[F^\prime_l+ G_l^\prime\,\pmb\sigma_1\pmb\sigma_2]\,\pmb\tau_1\pmb\tau_2\, P_l\biggl(\frac{\pmb k_1\pmb k_2}{k^2_F}\biggr),
\end{equation}
where $\pmb k_i$, $\pmb\sigma_i$, and $\pmb\tau_i$ are the nucleon momentum, spin, and isospin
operators,  and $N_0=2k_Fm^*/\pi^2\hbar^2$ with $k_F$ and $m^*$
denoting the Fermi momentum and nucleon effective mass,
respectively. For Skyrme interactions, all Landau parameters with $l>1$ are zero.
In the present study, we keep only $l=0$ terms in Eq.~\eqref{S_Landau_momentum},  and
in the coordinate representation  $H_\mathrm{ph}$ takes the form
\begin{equation}\label{S_Landau}
  H_\mathrm{ph}=N_0^{-1}[F^\prime_0+ G_0^\prime\pmb\sigma_1\pmb\sigma_2]\pmb\tau_1\pmb\tau_2\delta(\mathbf{r}_1-\mathbf{r}_2).
\end{equation}
The expressions for the Landau parameters
$F^\prime_0$, $G^\prime_0$ in terms of the Skyrme force parameters can be found in Ref.~\cite{Giai_PLB379}.   It was shown in Ref.~\cite{Giai_PRC57} that by approximating the full residual
interaction derived from a Skyrme force by its Landau-Migdal expansion truncated at the $l=0$ terms it is possible to
calculate accurately the isoscalar RPA modes and also to reproduce reasonably well the isovector RPA modes. Besides, as shown in~Ref.~\cite{Sever_JPhConf267}, 
for charge-changing GT and spin-dipole excitations, the results obtained in the Landau-Migdal $l=0$ approximation for $H_{ph}$ are close to the results obtained with the full residual interaction.

At $T\ne 0$, due to thermally unblocked configurations, the complexity of TQRPA calculations increases rapidly with the size of the configuration space.
The benefit of having a Landau-Migdal form~\eqref{S_Landau} is that it allows one
to construct a finite-rank separable particle-hole  interaction and thus to combine
the advantages of consistency and  simplicity~~\cite{Giai_PRC57}.
 This reduction enables us to transform the TQRPA equations into a  relatively simple secular equation of
low dimension and  perform TQRPA calculations in very large configurational space. The explicit form of the TQRPA secular equation for charge-changing transitions in hot nuclei is given 
in~Ref.~\cite{Dzhioev_PhAN79}. In what follows we will compare the results of TQRPA calculations with the Skyrme functionals with those performed by employing the QPM Hamiltonian~\cite{Dzhioev_PRC81}. To distinguish between the two approaches, we will refer to them as QPM-TQRPA and Skyrme-TQRPA.

\begin{table}
\caption{Landau parameters of infinite symmetric nuclear matter  calculated at density
$\rho = 0.16~(0.08)~\mathrm{fm}^{-3}$ using interactions SkM$^*$, SLy4, and SGII.}
\label{Landau}
\begin{tabular}{cccc}
\hline\noalign{\smallskip}
 &  $N^{-1}_0~(\mathrm{MeV\,fm^3})$ & $F'_0$ & $G'_0$  \\
\noalign{\smallskip}\hline\noalign{\smallskip}
SkM$^*$ &   194 (245) & 0.929 (1.133) & 0.316 (0.647)\\
SLy4    &   219 (276) & 0.818 (1.151) &$-0.138$ (0.365) \\
SGII    &  194 (244) & 0.726 (0.971) & 0.502 (0.740)\\
\noalign{\smallskip}\hline
\end{tabular}
\end{table}

In order to estimate the  sensitivity of our results with respect to the choice of Skyrme forces, three  different Skyrme  parametrizations, SLy4, SGII, and SkM*, are used in the present study.
The SLy4 parametrization~\cite{Chabanat_NPA635} is one of the most successful Skyrme forces and has been extensively used in recent years. The  force SGII~\cite{Giai_PLB379} has been 
successfully applied to study spin-isospin excitations in spherical and deformed nuclei, and  the SkM* interaction~\cite{Bartel_NPA79} is an example of the first-generation Skyrme 
parametrizations. In Table~\ref{Landau}, we give the Landau parameters $F'_0$ and $G'_0$ for symmetric nuclear matter at  density values $\rho=0.16,~0.08~\mathrm{fm}^{-3}$ for the Skyrme forces 
employed in the present study.
It is well known that existing Skyrme parametrizations predict the $G'_0$ value varying over a wide range (see, for example, Fig. 2 in Ref.~\cite{Niu_PRC85}).
The Skyrme parameter sets we use cover a great portion of this range and, therefore,  by employing these very different functionals,  we can estimate the theoretical uncertainty of 
the Skyrme-TQRPA approach.

Let us also make a short remark concerning the choice of the pairing interaction. As in~Ref.~\cite{Dzhioev_PRC81},  we employ in the present study the BCS Hamiltonian with a constant pairing strength.
The neutron and proton pairing strength parameters are fixed to reproduce the odd-even mass difference. At $T=0$, the resulting proton and neutron energy gaps are $\Delta_{p(n)}=1.57~(1.36)$~MeV 
for $^{56}$Fe and there are no pairing correlations (i.e., $\Delta_{p(n)}=0$) in $^{78}$Ni. Within the BCS approach, the phase transition in nuclei from the superfluid to normal state occurs
at critical temperatures $T_\mathrm{cr}\approx0.5\Delta$, where $\Delta$ is a pairing gap at $T=0$~\cite{Goodman_NPA352,Civitarese_NPA404}. Therefore, the inclusion of particle-particle
residual interactions does not affect the results for temperatures $T> T_\mathrm{cr}$.

Of course, the phase transition is a consequence of  the grand-canonical treatment which allows the particle number conservation only in average
and, strictly speaking, it occurs only in the thermodynamic limit. In finite systems, fluctuations around the mean field are important and they smooth the singularities associated with the phase 
transitions. As stated above, we use the grand-canonical description because nuclei in the stellar interior can exchange particles with the environment.
However, in a context other than the present one (e.g., in calculations of statistical properties of nuclei such as the level density or the specific heat), it is  necessary
to carry out a reduction to the canonical ensemble to restore the correct proton and neutron numbers. Such calculations were performed some times ago
by considering a many-body projected statistical density operator~\cite{Esebbag_NPA552}, by using the static-path approximation~\cite{Rossignoli_AnnPhys235}, and more recently 
by using a finite-temperature variation after projection BCS approach~\cite{Gambacurta_PRC88}.

\subsection{Iron isotope  $^{56}$Fe}\label{results_for_56Fe}

In Ref.~\cite{Dzhioev_PhAN79}, we have applied the Skyrme-TQRPA framework for analyzing thermal effects on the EC cross section  for $^{56}$Fe.
To simplify the consideration, it was assumed that EC on  $^{56}$Fe is dominated by allowed GT$_+$ transitions.
Considering the ground-state GT$_+$  strength distributions it was shown that QRPA calculations with the SkM*, SLy4, and SGII forces fairly well reproduce both the experimental~\cite{ElKateb_PRC49} 
and shell-model~\cite{Caurier_NPA653} centroid energies. The experimentally observed quenching of the total GT strength was reproduced in~Ref,~\cite{Dzhioev_PhAN79} by reducing the axial coupling 
constant $g_A$ by a quenching factor  $q=0.74$. The same quenching factor is used in the shell-model calculations of EC rates~\cite{Caurier_NPA653}. Of course, the QRPA calculations cannot reproduce 
all nuclear correlations needed to describe  the full GT resonance width  and its detailed structure. In this respect, the shell-model calculations  are clearly advantageous.
When considering  GT$_+$  strength distributions at $T\ne 0$, we have found that regardless of the Skyrme parametrization we use, the Brink-Axel hypothesis is violated within the TQRPA and the strength distribution evolves with temperature. Namely, thermal effects shift the GT$_+$ resonance  to lower energy and increase the contribution of low- and negative-energy transitions to the strength function.

\begin{figure}[t]
 \includegraphics[width=1.0\columnwidth]{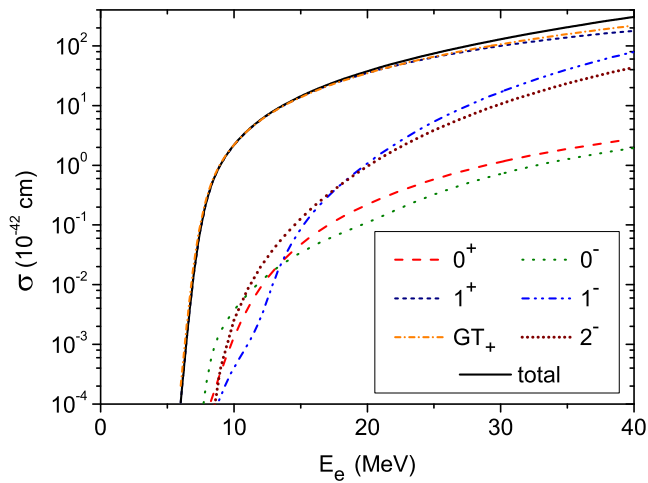}
\caption{\label{56Fe_CrSect_GrSt}  Electron-capture cross section on the ground state of $^{56}$Fe as a function of the incident electron energy $E_e$. 
The total cross section includes contributions of $J^\pi=0^\pm,1^\pm$, and $2^-$ multipole transitions calculated with the SkM* interaction. The dash-dotted line corresponds to 
 the $1^+$ contribution calculated with the GT$_+$ operator, i.e., neglecting momentum transfer.}
\end{figure}

\begin{figure*}[t]
\includegraphics[width=1.5\columnwidth]{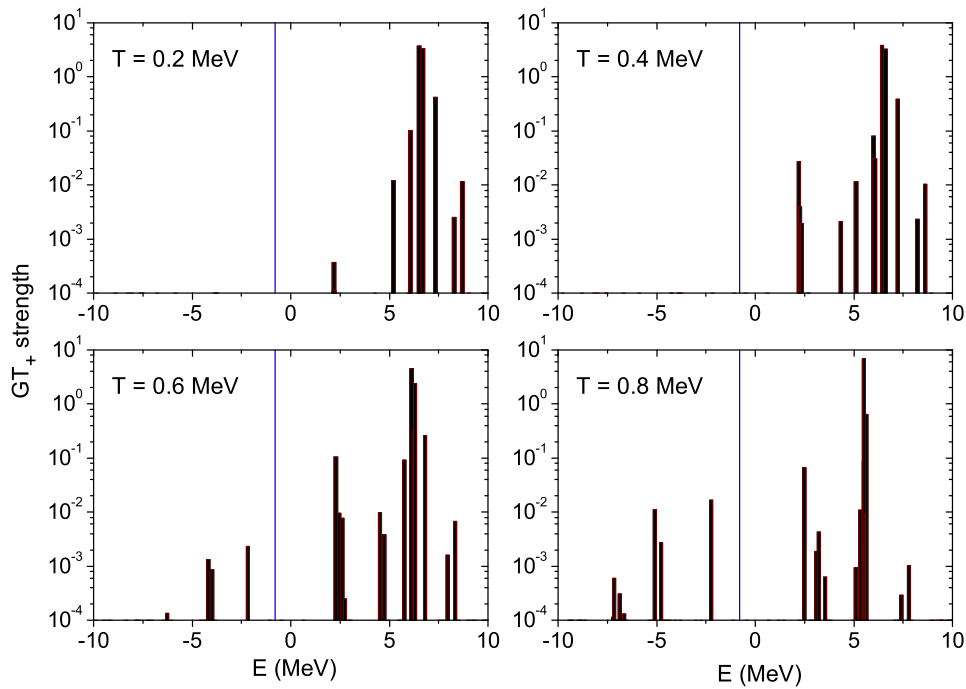}
\caption{\label{56Fe_GT} Temperature evolution of the GT$_+$ strength distribution in $^{56}$Fe vs transition energy.
 The strength distributions are obtained with the SkM$^*$  interaction for the bare operator GT$_+=\pmb\sigma t_+$.
 The solid vertical line indicates the value $\delta_{np}\approx -0.78$~MeV. The GT$_+$ strength below  $\delta_{np}$ corresponds to transitions to tilde one-phonon states. }
\end{figure*}

In the present study, we apply the Skyrme-TQRPA approach to compute EC rates  on $^{56}$Fe at conditions realised on the initial phase of core collapse and compare the results with those
obtained from QPM-TQRPA~\cite{Dzhioev_PRC81} and LSSM~\cite{Langanke_NPA673} calculations. To begin with, we first verify that within
the Skyrme-QRPA model GT$_+$ transitions dominate the EC reaction on $^{56}$Fe . In Fig.~\ref{56Fe_CrSect_GrSt}, we show the $0^\pm$, $1^\pm$, and $2^-$ multipole contributions to the 
ground-state EC cross section calculated with the SkM* interaction.
Note that the axial coupling constant $g_A$  is quenched for all the multipole excitations  by the factor $q=0.74$.
As seen from the figure,  the $1^+$ contribution completely dominates the cross section up to $E_e\approx 30$~MeV. Moreover,
by comparing the $1^+$ and GT$_+$ contributions, we observe that for energies $E_e < 30$~MeV the corrections due to final momentum transfer are negligible.
Therefore,  we conclude that at relatively low temperatures and densities (i.e., when the electron gas chemical potential $\mu_e\lesssim 10$~MeV), the electron capture on $^{56}$Fe
is determined by GT$_+$ transitions.

To make further discussion of EC rates more comprehensible, let us briefly recall the results of Refs.~\cite{Dzhioev_PRC81,Dzhioev_PRC92,Dzhioev_PhAN79} concerning thermal effects
on the GT$_+$ strength distribution in $^{56}$Fe.   For the sake of clarity, in Fig.~\ref{56Fe_GT} we display on a logarithmic scale the GT$_+$ distributions calculated at four different temperatures 
relevant to presupernova conditions. The strength distributions are obtained with the SkM$^*$ force. However, the observed thermal effects are qualitatively similar to those in calculations with 
the SGII and SLy4 forces as well as with the QPM Hamiltonian. Considering the temperature evolution of the upward ($E>0$)  GT$_+$ strength, we clearly see the violation of the Brink-Axel 
hypothesis within the TQRPA. Namely, thermal effects shift the resonance peak towards lower energies and enhance the strength of low-energy ($E\approx 2.5$~MeV) GT$_+$ transitions. 
As discussed in Ref.~\cite{Dzhioev_PRC81,Dzhioev_PRC92}, both the effects are caused by the thermal smearing of the proton and neutron Fermi surfaces, which leads to the vanishing of
 pairing correlations and  the softening of the proton-neutron repulsive residual interaction. Similar thermal effects on GT$_+$ distributions are predicted by the finite-temperature relativistic 
 RPA~\cite{Niu_PRC83} and  shell-model Monte-Carlo  calculations~\cite{Radha_PRC56}.

As evident from the figure, the downward ($E<0$) component of the GT$_+$ strength is affected by finite temperature as well. For $^{56}$Fe, this downward strength is located below
 $\delta_{np}\approx -0.78$~MeV and corresponds  to tilde-phonon states associated with negative-energy solutions of the TQRPA equations [see the discussion after Eq.~\eqref{phonon}].
In accordance with the detailed balance principle~\eqref{DBR2}, the temperature rise exponentially increases the transition strength to tilde-phonon states. We would like to stress again
that within the TQRPA framework the appearance of  $p\to n$ transitions  with $E<\delta_{np}$ stems from the doubling of the system degrees of freedom within the TFD.

To reveal the importance of tilde-phonon states, in~Ref.~\cite{Dzhioev_PhAN79} the TQRPA
EC cross sections for $^{56}$Fe were compared with those obtained by the finite-temperature RPA (FTRPA) calculations~\cite{Fantina_PRC86} using the same Skyrme parametrizations.
Like the hybrid model, the FTRPA
 neglects negative-energy solution of RPA equations and therefore misses GT$_+$ strength which contributes to an exoergic electron capture process. It was shown in~Ref.~\cite{Dzhioev_PhAN79}
that if we neglect exoergic GT$_+$ transitions within the TQRPA, both the approaches produce rather close EC cross sections  which drop to zero as the electron energy falls below some
minimum threshold value. However, as soon as GT$_+$ exoergic transitions to tilde-phonon states are taken into account, the reaction threshold disappears and EC becomes possible for arbitrary small incident electron energy.

\begin{figure}[t]
 \includegraphics[width=1.0\columnwidth]{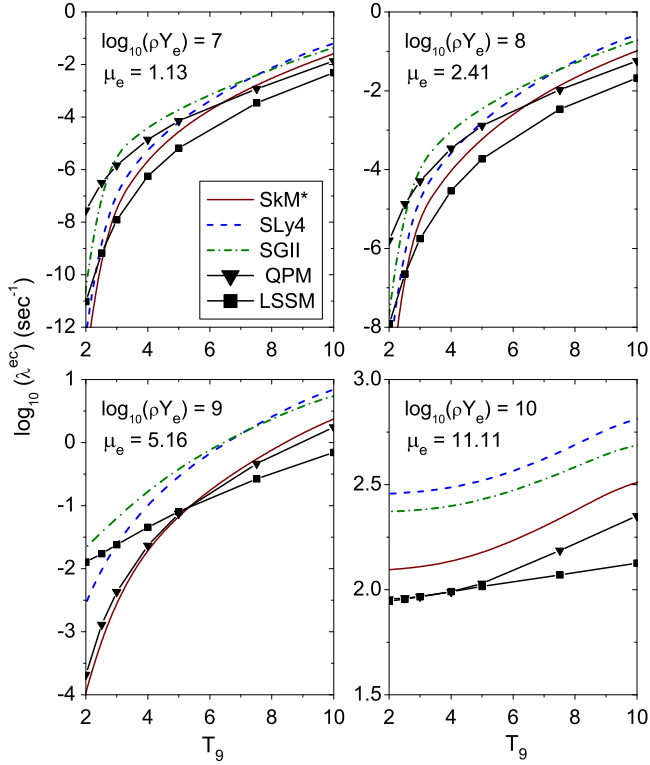}
\caption{ \label{56Fe_ECrates_Temp}  Electron capture rates for $^{56}$Fe  as functions  of the temperature $T_9$ ($T_9=10^9\,\mathrm{K}\approx 0.086\,\mathrm{MeV}$) at
 selected densities $\rho Y_e$ (g\,cm$^{-3}$). $\mu_e$ is the electron gas chemical potential (including the rest mass) in MeV calculated for a given density and temperature $T_9=2$.   
 On each panel, the TQRPA results obtained  with the  SkM$^*$, SLy4, and SGII interactions are compared with the LSSM rates. The rates obtained in~Ref.~\cite{Dzhioev_PRC81} by employing a phenomenological QPM Hamiltonian are also shown. }
 \end{figure}

\begin{figure}[t]
\includegraphics[width=1.0\columnwidth]{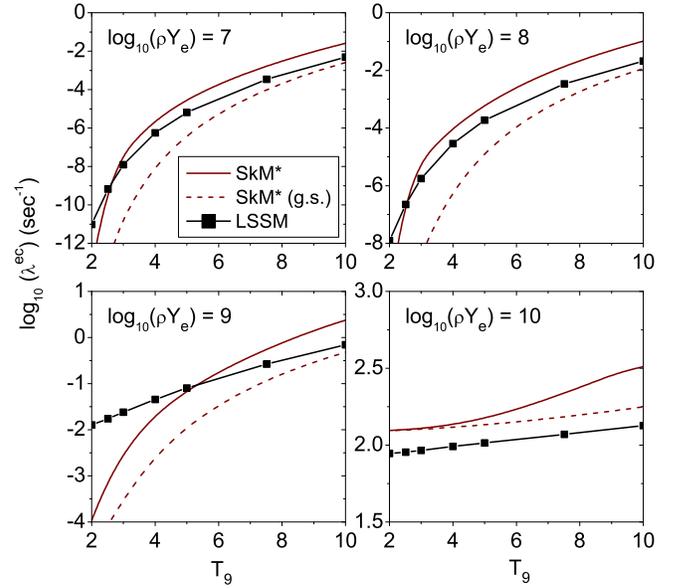}
\caption{\label{56Fe_ECrates_Temp_2} Electron capture rates for $^{56}$Fe computed with and without taking into account thermal effects on the GT$_+$ strength. The results obtained
with the  SkM$^*$ interaction are shown along with the LSSM rates.}
\end{figure}

Now we present the Skyrme-TQRPA results for stellar electron capture rates on $^{56}$Fe and compare them with the previous theoretical calculations. The rates  we present include
the GT$_+$ contribution only and they are computed according to Eq.~\eqref{rate_GT}.
In Fig.~\ref{56Fe_ECrates_Temp}, the variation of electron capture rates with temperature and density is displayed
for the Skyrme forces SLy4, SGII, and SkM$^*$. Note that $T_9$ gives stellar temperature in units of $T_9=10^9$~K ($1~\mathrm{MeV}\approx 11.6~T_9$).
As seen from the figure, the rates obtained with the different Skyrme parametrizations reveal similar trends.
Namely, the rates increase with temperature and  density, and the thermal enhancement  is most significant at low densities, when
the electron chemical potential is small. In our model, two factors contribute to the thermal enhancement of EC rates at low densities: First, temperature rise increases the
number of high-energy electrons in the tail of the Fermi-Dirac distribution, which then excite the GT$_+$ resonance. Second, as discussed earlier, finite
temperature allows low- and negative-energy GT$_+$ transitions which dominate the rate  if the electron capture into the resonance state is energetically suppressed.
To examine qualitatively how thermal changes of the GT$_+$ strength affect the EC rates on $^{56}$Fe,  we compare in Fig.~\ref{56Fe_ECrates_Temp_2} the rates
with those obtained assuming that EC proceeds on the nuclear ground state. Referring to the figure, at low densities thermal effects on the GT$_+$ strength function
 enhance the rate by two orders of magnitude. In  contrast, thermal effects are less important at high densities when
 the electron gas chemical potential becomes large enough so that  transitions to the GT$_+$ resonance dominate EC. At such
conditions, the capture rates are mainly sensitive to the total GT$_+$ strength and its centroid.
Remember, however,  that the  GT$_+$ resonance shifts to lower energies with temperature.
This explains why even at $\log_{10}(\rho Y_e)=10$ when $\mu_e\approx 10$~MeV the rate slightly increases with temperature.

Though the general behavior of the rates as a function of  temperature and density is the same, Fig.~\ref{56Fe_ECrates_Temp} indicates that the spread of the values obtained with
the different Skyrme forces can reach two orders of magnitude.  The discrepancy is most pronounced at low temperatures and densities when
the rates are highly sensitive to the GT$_+$ distribution details. According to our TQRPA calculations, among the considered Skyrme functionals, the one based on the SkM$^*$ interaction
predicts the  GT$_+$  distribution slightly shifted to higher energies (see Fig.~2 in Ref.~\cite{Dzhioev_PhAN79}).  For this reason, the rates obtained with the SkM$^*$ force
are generally smaller than those computed with  the SLy4 or SGII forces. This observation is in line with~\cite{Fantina_PRC86}, where the SkM* parametrization,
along with others, was used to compute EC rates on $^{56}$Fe.

Figure~\ref{56Fe_ECrates_Temp} also compares the Skyrme-TQRPA rates to those obtained in the LSSM~\cite{Langanke_NPA673} and QPM-TQRPA approaches~\cite{Dzhioev_PRC81}.
 As  seen from the bottom-right panel of the figure, there is an excellent agreement between the QPM-TQRPA~ and the shell-model results
at high densities and low temperatures when the rates are dominated by the ground-state GT$_+$ resonance contribution. This agreement is not surprising, since the
QPM Hamiltonian parameters were adjusted to available experimental data on the GT$_+$ distribution  in $^{56}$Fe~\cite{ElKateb_PRC49}.  At the same conditions, the Skyrme-TQRPA calculations 
yield somewhat higher rates. There are two  reasons for this discrepancy: (1)  As follows from our calculations (see Fig.~1 in~Ref.~\cite{Dzhioev_PhAN79}), even
with the quenching factor $q=0.74$, the  total GT$_+$ strength in $^{56}$Fe  obtained with the Skyrme interactions SkM$^*$, SGII, and SLy4 somewhat overestimates the experimental
and the shell-model values. The largest total GT$_+$ strength is obtained with the SLy4 parametrization and it overestimates the shell-model and QPM results by a factor of 2.  
(2) Besides, the GT$_+$ resonance peaks calculated with the Skyrme-QRPA are slightly shifted to lower energies with  respect to the shell-model ones.  The discrepancy between the Skyrme-TQRPA and LSSM 
rates at $\log_{10}(\rho Y_e)=10$  even increases with temperature, as the GT$_+$ resonance undergoes the temperature induced downshift.

At lower densities, when the electron gas chemical potential $\mu_e$ is smaller or comparable with the GT$_+$ resonance energy, the EC rates are very sensitive to details of the GT$_+$ distribution.
As shown in the top panels of Fig.~\ref{56Fe_ECrates_Temp}, the TQRPA rates generally overestimate the shell-model values for densities when $\mu_e$ is smaller than the reaction $Q$ value
($Q=M(^{56}\mathrm{Mn})-M(^{56}\mathrm{Fe})=4.2$~MeV).  The detailed analysis performed in~Ref.~\cite{Dzhioev_PRC81} showed that this disagreement stemmed from larger strength of thermally
unblocked low- and negative-energy transitions predicted by the TQRPA model. This conclusion is also supported by Fig.~\ref{56Fe_ECrates_Temp_2}, showing that the inclusion of thermal effects 
on the GT$_+$ strength function makes the TQRPA rates higher than the LSSM ones at low densities.  For the moderate density $\log_{10}(\rho Y_e)=9$, $\mu_e\approx Q$ and the near-threshold
part of the GT$_+$ strength dominates the rates at low temperatures. The TQRPA calculations with the SkM$^*$ force as well as with the QPM Hamiltonian do not reproduce this 
strength and, therefore, the respective rates appear to be suppressed with respect to the LSSM ones. In contrast, due to the displacement of the GT$_+$ resonance to lower energies, calculations
with the SGII and SLy4 interactions lead to more enhanced rates. As was shown in~Ref.~\cite{Dzhioev_PRC81}, the disagreement between LSSM and TQRPA rates at moderate densities and low temperatures 
can be removed by considering the fragmentation of the GT$_+$ strength.

\subsection{Neutron-rich  isotope $^{78}$Ni}

Now we turn our discussion to  the  neutron-rich nucleus $^{78}$Ni.  In Ref.~\cite{Sullivan_APJ816}, a complete study was performed on the sensitivity
of the core-collapse dynamics to variations of EC rates in medium-heavy nuclei. In particular, it was shown  that core-collapse supernova simulations
are most sensitive to individual EC rates on neutron-rich nuclei in the upper $pf$ and $pfg/sdg$ shells. Specifically, it was found that nuclei  around $N{=}50$ closed neutron shell at and above 
$^{78}$Ni have the largest impact on the mass of the inner core at bounce and on the peak neutrino luminosity. In the present work, we compute EC rates for  $^{78}$Ni and compare them with an analytic 
approximation widely used in core-collapse simulations.

Because of large neutron excess the ground-state threshold energy for the electron capture on  $^{78}$Ni, $Q = {M}(^{78}\mathrm{Co})-{M}(^{78}\mathrm{Ni})\approx 20.7$~MeV~\cite{Moller_ADNDT109} is 
significantly enhanced when  compared to $^{56}$Fe. Hartree-Fock calculations for $^{78}$Ni with the Skyrme interactions SkM$^*$, SGII, and SLy4 predict closed $1f_{7/2}$ proton and $1g_{9/2}$ neutron 
subshells.
For neutron-rich nuclei with $N >40$ and $Z < 40$, the independent particle shell-model predicts that at $T=0$ all allowed GT$_+$ transitions of valence protons are Pauli blocked, owing to the complete 
occupation of the $pf$ neutron orbitals. It was first demonstrated in Ref.~\cite{Cooperstein_NPA420}   that the electron capture on nuclei with protons in the $pf$ shell and $N > 40$ could compete with 
the capture on free protons if first-forbidden (i.e., $0^-,~1^-$, and $2^-$) transitions are taken into account in addition to thermally unblocked  GT$_+$ ones.
In Ref.~\cite{Langanke_PRC63}, an alternative unblocking mechanism --- configuration mixing induced by the residual interaction --- was considered  by employing the hybrid model and it was found that 
unblocking effect was not very sensitive to increasing temperature.

In Ref.~\cite{Dzhioev_PRC81}, using $^{76,78,80}$Ge as examples, a detailed analysis was performed of thermal effects on  GT$_+$  transitions in neutron-rich nuclei within the TQRPA.   
It was shown that both thermal excitations and the configuration mixing due to pairing correlations promote protons to the $sdg$ shell and remove neutrons from the $pf$ orbitals. As a result, 
the particle-particle $1g^p_{9/2}\to 1g^n_{7/2}$ and hole-hole $1f^p_{7/2}\to 1f^n_{5/2}$ proton-to-neutron transitions become unblocked and dominate the GT$_+$ contribution to EC on neutron-rich nuclei
with $N >40$ and $Z < 40$. However, in contrast to the hybrid model, the TQRPA predicts that both the strength and the energy of unblocked GT$_+$ transitions are temperature dependent due 
to the destructive interference of thermal excitations and pairing correlations. As a result,  the unblocking effect for neutron-rich nuclei turns out to be quite sensitive to temperature rise.

\begin{figure*}[t]
\includegraphics[width=1.5\columnwidth]{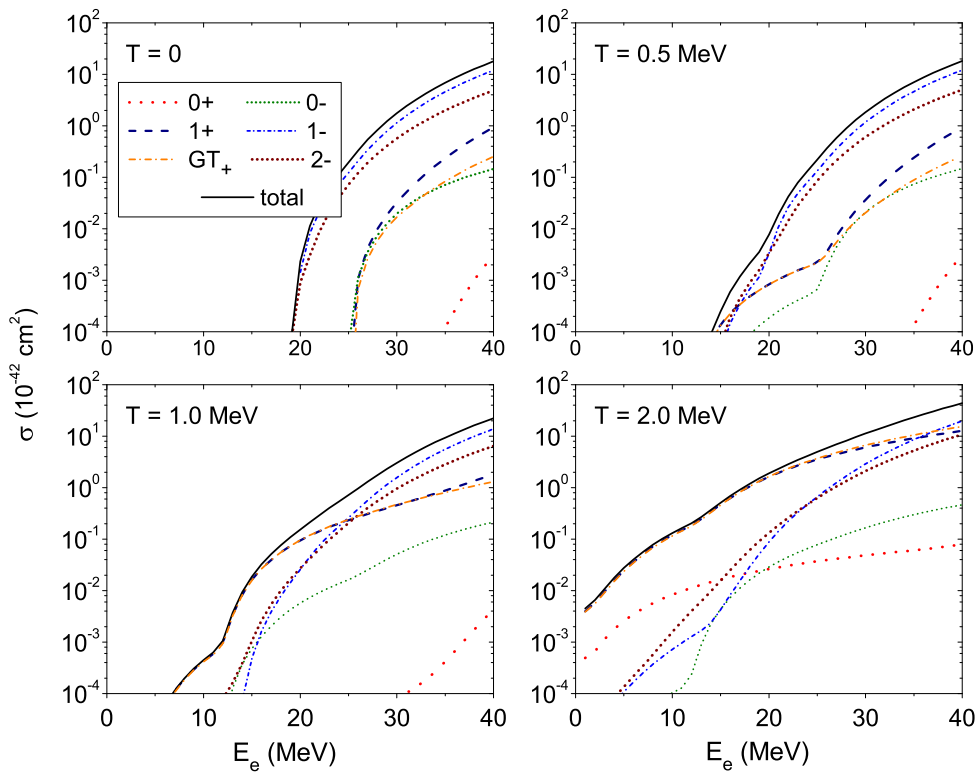}
\caption{ \label{78Ni_CrSect_Temp_SkM} Electron capture cross sections on $^{78}$Ni at different temperatures $T$. The cross sections are obtained using the SkM$^*$ interaction. }
\end{figure*}

To demonstrate the unblocking effect for $^{78}$Ni, in Fig.~\ref{78Ni_CrSect_Temp_SkM} we compare the ground-state ($T=0$) EC cross sections with those calculated at three $T$ values relevant to
core collapse. The cross sections depicted in the figure are computed according to Eq.~\eqref{sigmaET} with the SkM$^*$ interaction. They include the contributions from 
$0^\pm,1^\pm$, and $2^-$ multipole transitions. According to our RPA calculations with  SkM$^*$, SLy4, and SGII forces, the ground-state cross section for EC on $^{78}$Ni is mediated by $1^-$ and $2^-$ 
forbidden transitions, while allowed $1^+$  transitions are almost completely blocked and their nonvanishing contribution  is a combined effect  of nonorthogonal proton and neutron wave functions and 
finite-momentum transfer. Referring to Fig.~\ref{78Ni_CrSect_Temp_SkM}, a strong temperature  dependence of the cross section at energies below the ground-state reaction threshold (i.e., $E_e<20$~MeV) 
is mainly due to increasing contribution of $1^+$ transitions. On the scale used in the figure, this effect manifests itself as a gradual reduction of the energy gap with temperature (compare the panel 
for $T=0$ with the panels for $T=0.5$ and $1.0$ MeV) and  at $T=2.0$ the gap disappears at all. Moreover, as the temperature increases,  the energy range $E_e$ of incoming electrons increases, where $1^+$
transitions dominate the cross section. Although the contribution of forbidden transitions  also increases with temperature (see the discussion below),  comparing  the multipole decompositions of
the ground-state cross section with those obtained at $T\ne 0$,  we conclude that the  cross-section enhancement at finite temperatures is essentially due to thermal effects on $1^+$ transitions.

\begin{figure}[t]
\includegraphics[width=1.0\columnwidth]{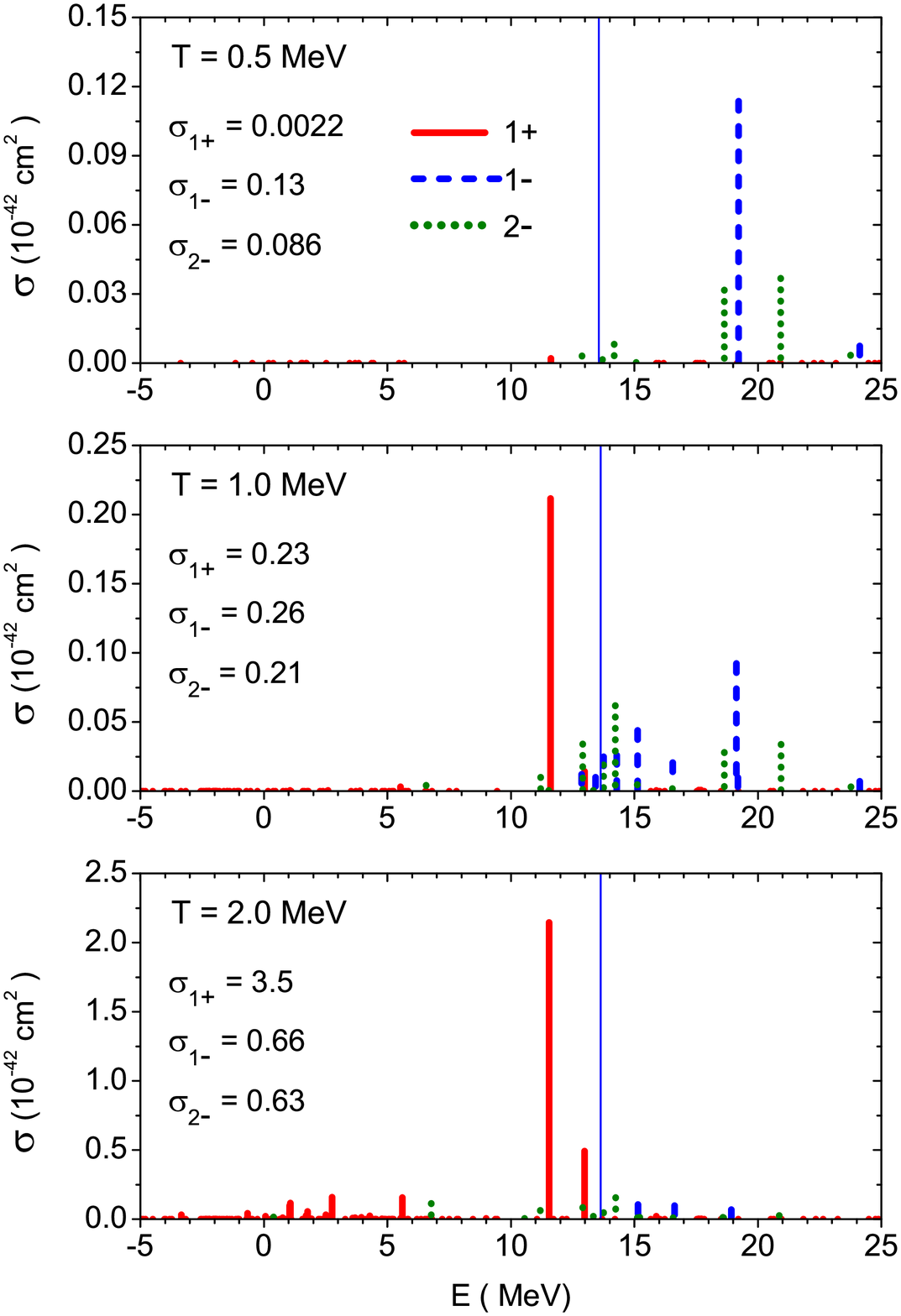}
\caption{\label{78Ni_DiffCrSect}   Leading multipole contributions $J^\pi = 1^+,~1^-$, and $2^-$ to the differential capture cross section of 25-MeV electrons on $^{78}$Ni.
 The calculations are performed  for  $T=0.5,~1.0$, and $2.0$~MeV using   the SkM$^*$ interaction. Total multipole contributions $\sigma_{J^\pi}$ (in $10^{-42}~\mathrm{cm}^{2}$) are shown for
  each temperature. The solid vertical line indicates the value $\delta_{np}$. For $E<\delta_{np}$, the differential cross section is due to transitions to tilde one-phonon states. }
\end{figure}

To explain the temperature evolution of  the $J^\pi =1^+,~1^-,$ and $2^-$ leading multipole components, in Fig.~\ref{78Ni_DiffCrSect} we show the EC differential cross sections computed 
for incoming electrons with $E_e=25$~MeV. For  25-MeV electrons, three different situations are realised, depending on temperature (see Fig.~\ref{78Ni_CrSect_Temp_SkM}):
(i) at $T=0.5$~MeV the $1^+$ contribution  is small and $1^-$, $2^-$ forbidden transitions dominate the cross section; (ii) at  $T=1.0$~MeV the allowed and forbidden components are of the same order; 
and (iii) at $T=2.0$~MeV the cross section is dominated by the thermally unblocked  allowed $1^+$ transitions. Notice also that for 25-MeV electrons and $T\ge 0.5$~MeV, we can neglect
the momentum dependence of the $1^+$ multipole operators and consider them in the long wavelength limit. This observation is verified in Fig.~\ref{78Ni_CrSect_Temp_SkM} where we compare the $1^+$ 
contributions computed  by taking into account the full $q$ dependence of the transition operators with those obtained with the GT$_+$ operator.

For the double magic nucleus  $^{78}$Ni, only thermal effects are responsible for the unblocking of GT$_+$ transitions; therefore, at low temperatures the $1^+$ component of the cross section
is negligibly small, as shown in~Fig.~\ref{78Ni_DiffCrSect}.  With increasing temperature,   the  GT$_+$ transition $1f^p_{7/2}\to 1f^n_{5/2}$ with energy $E\approx 11.5$~MeV becomes unblocked and 
it dominates the $1^+$ differential cross section at $E_e=25$~MeV. The  $1g^p_{9/2}\to 1g^n_{7/2}$ transition with energy $E\approx 13.1$~MeV  is also thermally unblocked in $^{78}$Ni but its 
contribution is smaller.
It should also be emphasized that within the TQRPA both the thermally unblocked transitions correspond to tilde-phonon states, i.e., to negative-energy solutions  of the TQRPA equations.
Hence, such transitions would not be taken into account without the doubling the Hilbert space.  As evident from Fig.~\ref{78Ni_DiffCrSect}, besides the considered strong or resonance transitions, 
thermal effects unblock some other $1^+$ low- and negative-energy  transitions. At $E_e=25$~MeV, their contribution to the cross section is negligible, but
it is these nonresonance transitions that remove the reaction threshold and dominate the cross section for $E_e<11$~MeV electrons.

\begin{table}
\caption{The total  Gamow-Teller strengths $B(\mathrm{GT}_+)$ calculated within the TQRPA with the different Skyrme forces.
The unperturbed  mean-field strengths  are given in parentheses. Note that  $B(\mathrm{GT}_+)$ values
 are obtained with the bare operator $\mathrm{GT}_+=\pmb\sigma t_+$.  }
\label{table2}
\begin{tabular}{cccccc}
\hline\noalign{\smallskip}
$T$ (MeV)   & 0.5 & 1.0 & 1.5 & 2.0  \\
\noalign{\smallskip}\hline\noalign{\smallskip}
SkM$^*$ & 0.13 (0.22) & 0.18 (0.33) & 0.40 (0.83) & 0.81 (1.8)  \\
SLy4    & 0.11 (0.17) & 0.13 (0.19) & 0.32 (0.39) & 0.91 (1.0) \\
SGII    & 0.14 (0.23) & 0.15 (0.24) & 0.20 (0.46) & 0.38 (1.1) \\
\noalign{\smallskip}\hline
\end{tabular}
\end{table}

In Table~\ref{table2}, we compare the total thermally unblocked strengths $B(\mathrm{GT}_+)$ in $^{78}$Ni calculated with the different Skyrme parametrizations. The table
 shows the result obtained with and without taking into account RPA correlations.  As the data of the table suggest, the strength of thermally unblocked GT$_+$ transitions varies  depending
 on the Skyrme parametrization used.  To explain this, we note that in the absence of pairing, the unperturbed mean-field strength of thermally unblocked $j_p\to j_n$ single-particle transition is 
 proportional to $n_{j_p}(1-n_{j_n})$, where $n_{j}$ are the occupation numbers for single-particle orbitals  (see Ref.~\cite{Dzhioev_PRC81} for more details).
 Therefore, variation of the unperturbed GT$_+$ strength is mainly due to difference in the occupation numbers of the proton $1f^p_{7/2}$, $1g^p_{9/2}$
 and neutron $1f^n_{5/2}$, $1g^n_{7/2}$  orbitals in $^{78}$Ni,  predicted by the SkM$^*$, SGII, and SLy4 effective interactions. In their turn, the occupation numbers depend on the single-particle
 Hartree-Fock energies.  Besides, the particle-hole correlations reduce the GT$_+$ strength.
As the SGII force assumes the largest particle-hole correlations due to spin-isospin interaction
(since it has the largest $G'_0$ values; see Table~\ref{Landau}), it produces the strongest reduction of the GT$_+$ strength.

Let us now briefly analyze thermal effects on the forbidden $1^-$ and $2^-$ components of the EC cross section.  According to our calculations, at zero and low temperatures  the  
$1^-$ and $2^-$  components of the cross section are determined by  particle-hole resonance transitions from the closed proton orbital $1f^p_{7/2}$. Namely,  $1f^p_{7/2}\to 2d^n_{3/2}$ 
and $1f^p_{7/2}\to 1g^n_{7/2}$  dominate the $2^-$ contribution, while    $1f^p_{7/2}\to 2d^n_{5/2}$ dominates the $1^-$ contribution. Thermal effects deplete the $1f^p_{7/2}$ orbital and 
promote protons to higher orbitals. Besides, the neutron subshell $1g_{9/2}$ becomes
partially unblocked.  As these take place, low-energy $p\to n$ transitions become possible, which also contribute to thermal enhancement of the EC cross section. They are 
$2^-$ transitions $1f^p_{7/2}\to 1g^n_{9/2}$ ($E\approx 14.2$~MeV), $1f^p_{5/2}\to 1g^n_{9/2}$ ($E\approx 6.8$~MeV), and $1^-$ transitions $2p^p_{3/2}\to 2d^n_{3/2}$ ($E\approx 15.2$~MeV), 
$1f^p_{5/2}\to 1g^n_{7/2}$ ($E\approx 16.7$MeV). Because of the increased phase space,   the contribution of thermally unblocked low-energy $1^-$ and $2^-$ transitions to the cross section can be 
larger than that of the resonance transitions. This is clearly seen in  Fig.~\ref{78Ni_DiffCrSect}, showing that at $E_e=25$~MeV and high temperatures the $1^-$ and $2^-$ components of the
 differential cross section are dominated by the unblocked transitions.  Referring to Fig.~\ref{78Ni_CrSect_Temp_SkM},   even for 40~MeV electrons the role of thermally unblocked $1^-$ and $2^-$ 
 transitions is substantial and their contribution along with that  of $1^+$  unblocked transitions enhances the cross section at $T=2.0$~MeV by more than a factor of 2 with respect to the ground-state
  value. It is noteworthy  that the increasing role of thermally unblocked $1^\pm$ and $2^-$ transitions means that as temperature goes higher the  EC process on $^{78}$Ni will produce a larger fraction 
  of high-energy neutrinos.

\begin{figure}[t]
 \begin{centering}
\resizebox{0.5\textwidth}{!}{ \includegraphics{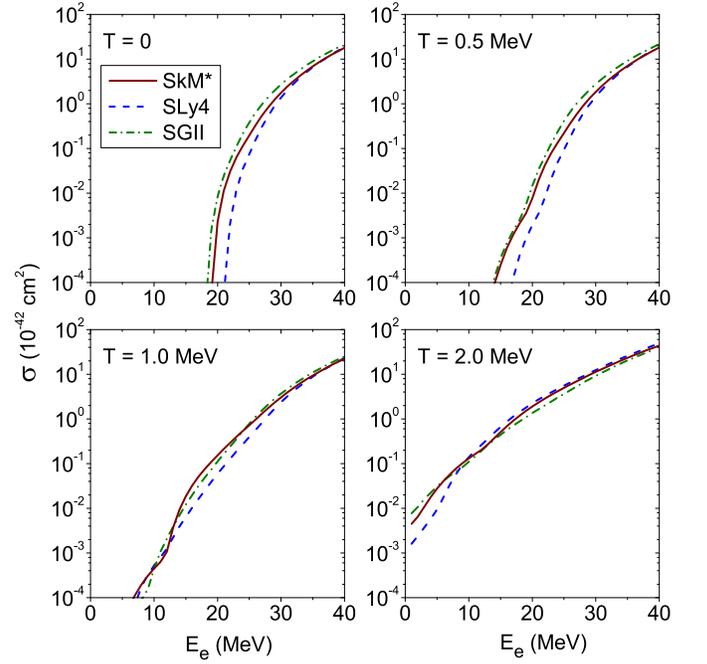} }
\caption{ Electron capture cross sections for $^{78}$Ni at four temperatures. The results are obtained by employing  the TQRPA calculations with the SkM$^*$,  SLy4, and SGII interactions. 
 For the SkM$^*$,  SGII, and SLy4 forces, the energy gaps in the cross section at $T=0$ are 19.1, 18.2, and 21.5 MeV, respectively. }
 \label{78Ni_CrSect_Temp}
 \end{centering}
\end{figure}

To show that the thermal effects discussed above are rather insensitive to the choice of the Skyrme interaction, in Fig.~\ref{78Ni_CrSect_Temp} we compare the EC cross sections for $^{78}$Ni computed
with the SkM$^*$, SGII, and SLy4 parametrizations. As seen from the figure, all the Skyrme forces  predict rather close values for the reaction threshold at $T=0$. Regardless of the Skyrme force used, 
the contribution of thermally populated states removes the threshold energy and significantly enhances the low-energy cross section. Thermal effects become less pronounced for high-energy electrons. 
However, even at $E_e=40$~MeV they increase the cross section by more than a factor of 2 when the temperature reaches 2.0~MeV. We also see from the figure that with increasing electron energy and 
temperature the spread  in the cross sections is reduced.

\begin{figure}[t]
 \begin{centering}
\resizebox{0.5\textwidth}{!}{ \includegraphics{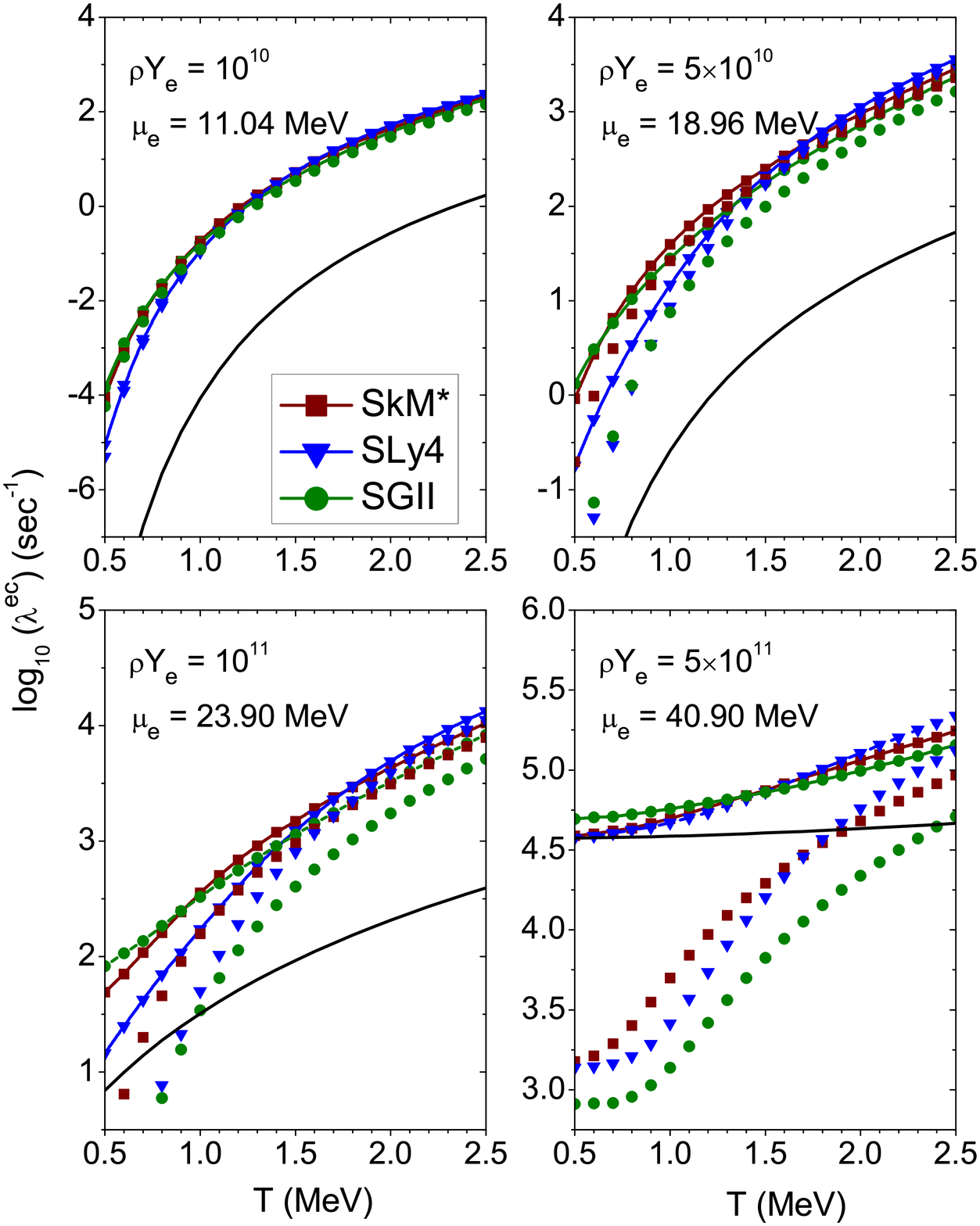} }
\caption{ Electron capture rates for $^{78}$Ni  calculated at selected densities $\rho Y_e$ (g\,cm$^{-3}$) as functions  of temperature. Each type of symbol corresponds to a specific Skyrme 
parametrization. Unconnected symbols represent the   contribution  of $1^+$ transitions to the respective rate. The solid line is the EC rate calculated according to parametrization~\eqref{EC_param}. 
$\mu_e$ is the electron gas chemical potential (including the rest mass) calculated for a given density and temperature $T=0.5$~MeV.}
 \label{78Ni_ECrates_Temp}
 \end{centering}
\end{figure}

Figure~\ref{78Ni_ECrates_Temp} compares the EC rates for $^{78}$Ni as obtained in the TQRPA model with different Skyrme sets by integrating the cross section [see Eq.~\eqref{rate_CrSec}]. 
The rates include the contribution of $0^\pm, 1^\pm$, and $2^-$ transitions.
In each panel,  we also show the individual contribution of allowed $1^+$ transitions to the rate. As expected, the rates increase with  temperature and density but they are
almost temperature independent at high densities.  Referring to  the top-left panel in the figure, at low densities the EC rate is determined by $1^+$ transitions. This can be understood as follows. 
For the density $\rho Y_e = 10^{10}~\mathrm{g\, cm^{-3}}$ the chemical potential is $\mu_e\approx 11$~MeV and, as seen in Fig.~\ref{78Ni_DiffCrSect},
this energy is not enough for both resonance and thermally unblocked $1^-$ and $2^-$   transitions. At the same time, this energy is sufficient for thermally unblocked Gamow-Teller  transitions which 
dominate the rate. At higher densities the chemical potential is large enough for the strong resonance $1^-$ and $2^-$ transitions and they dominate the rate at low temperatures. However, the role of 
allowed $1^+$ transitions increases with temperature and at $T>2.0$ MeV their contribution is comparable or even higher than that of forbidden transitions.
Comparing the results obtained with different Skyrme forces, we can see that the spreading of the capture rates is less than one order of magnitude.
As for EC rates on $^{56}$Fe, the spreading is most significant at low temperatures and densities and it  reduces with $T$ and $\rho$.

In~\cite{Fuller_APJ293}, Fuller \textit{ et al.} proposed an analytic approximation for electron capture rates which is based on the $Q$-value dependence of the capture rate. For neutron-rich nuclei
this approximation was later parameterized by Langanke \textit{et al}.~\cite{Langanke_PRL90} to fit their detailed microscopic calculations. It can be written as
\begin{equation}\label{EC_param}
  \lambda = \frac{\ln 2\cdot B}{K}\Bigl(\frac{T}{m_e c^2}\Bigr)^5\bigl[ F_4(\eta) - 2\chi  F_3(\eta)  + \chi^2 F_2(\eta)\bigr],
\end{equation}
where $F_k$ are the Fermi integrals of rank $k$ and degeneracy $\eta$, $\chi = -(Q+\Delta E)/T$~\footnote{Note that in our definition $Q=M_f - M_i$, while in~Ref.~\cite{Langanke_PRL90} 
the $Q$ value is defined with opposite sign.}, and $\eta =\chi + \mu_e/T$. The fit parameters $B$ and $\Delta E$ represent, respectively, effective values for the transition strength
(Gamow-Teller plus forbidden) and energy difference between the final and initial excited states. With the values $B=4.6$ and $\Delta E = 2.5$~MeV, a good agreement is achieved between 
the parametrization~\eqref{EC_param} and microscopic calculations for a very large number of nuclei~\cite{Langanke_PRL90}. In Fig.~\ref{78Ni_ECrates_Temp}, we compare the rates evaluated
 with this approximation and those from the present Skyrme-TQRPA calculations. As can be seen in the bottom-right panel of the figure, there is a good agreement between both calculations at 
 high densities and low temperatures,  when the rate is mainly determined by the total strength of $1^-$ and $2^-$ forbidden transitions from the  parent ground state. This means that the present 
 Skyrme-RPA calculations at $T=0$ predict the strength of forbidden transitions rather close to the value used in~Eq.~\eqref{EC_param}. However, in the present model, the thermal effects unblock some
  additional strength of $p\to n$ transitions. This makes the TQRPA rates at $\rho Y_e = 5\times10^{11}~\mathrm{g\, cm^{-3}}$ more sensitive to temperature rise than predicted by the 
  approximation~\eqref{EC_param}.  The results of two approaches differ significantly at lower densities when the electron chemical potential is comparable or  below the reaction $Q$ value  
  (i.e., $\mu_e \lesssim 20$~MeV). Under such  conditions, the rates are sensitive to the fragmentation of the $p\to n$ transition strength. Since
Eq.~\eqref{EC_param} is based on the assumption that  all the strength is concentrated in a single state above the $Q$ value, the resulting rates are much smaller than those predicted by the TQRPA.

\section{Conclusion}\label{conclusion}

In this work, we have further developed and generalized a  theoretical framework for modeling the process of electron capture on nuclei in presupernova conditions.
By applying the Walecka-Donnelly formalism, we expressed the cross section for the electron capture on a hot nucleus  through the temperature- and momentum-dependent spectral
functions for charge, longitudinal, transverse electric, and transverse magnetic  charge-changing operators.  The method we employed to compute spectral functions
is a self-consistent proton-neutron QRPA with the Skyrme interaction extended to finite temperatures by the thermo-field-dynamics formalism.
It is shown that the method is thermodynamically consistent since it preserves the detailed balance principle for $p\to n$ and $n\to p$ spectral functions.

Choosing $^{56}$Fe and $^{78}$Ni as examples, we have performed  illustrative electron capture calculations
for the iron-group and neutron-rich nuclei. For $^{56}$Fe, our present calculations
with three different Skyrme parametrizations reveal the same thermal effects on the GT$_+$ strength function as
those found in the previous study based on the QPM Hamiltonian.  In particular, increasing  temperature shifts the GT$_+$ resonance to lower energies and makes  low- and negative-energy GT$_+$ 
transitions possible.
For $^{78}$Ni, we have found that thermal effects unblock both GT$_+$  and  low-energy first-forbidden transitions, thereby increasing the  EC cross section
significantly.  It is interesting to note that for $^{78}$Ni different Skyrme forces predict finite-temperature cross sections which do not differ significantly. This result is in line with the 
conclusion made in~Ref.~\cite{Dzhioev_PhAN72} when considering the EC cross sections for $^{56}$Fe.

Electron capture rates have been calculated for different densities and temperatures of stellar matter and then compared with the results of other approaches.
For $^{56}$Fe, the Skyrme-TQRPA rates reproduce the temperature dependence of the rates predicted by shell-model calculations, but at low and high densities our results are generally
larger. The observed discrepancy is mainly due to violation of the Brink-Axel hypothesis within the TQRPA, which leads to a larger strength of thermally
unblocked transitions. On the other hand, the underestimation of the EC rate  at temperatures and  densities when  $\mu_e\approx Q$ indicates the importance of multinucleon
correlations beyond the RPA.  For neutron-rich $^{78}$Ni, the role of the thermal effects on the EC rates is even more substantial.
It is shown that the inclusion of thermally unblocked GT$_+$ and first-forbidden $1^-$ and $2^-$ transitions significantly enhances the rates in comparison with the values predicted
by the parametrization~\eqref{EC_param}.

In the present application, correlations described by the TQRPA have been taken into account. Whereas much of the essential physics is already captured by the model, the detailed comparison
 with the shell-model results indicates that the approach should be further improved. In particular, the fragmentation of the GT$_+$ strength plays a significant role at low temperature and 
 densities of the supernova environment. Therefore, a further improvement of the model is to go beyond the TQRPA and take into account higher order correlations. For the finite-rank separable 
 residual approximation for the Skyrme interaction this can be done by coupling the thermal phonon states with more complex (e.g., two-phonon) configurations. For charge-exchange excitations
  at zero temperature, the phonon coupling was considered within the QPM model~\cite{Kuzmin_JPG10} and  with the self-consistent Skyrme-based calculations~\cite{Severyukhin_RJPh58}.  Most 
  recently, the fragmentation of the finite-temperature Gamow-Teller strength due to the particle-vibration coupling was studied within the  relativistic time-blocking 
  approximation~\cite{Litvinova_arxiv2019}. In particular, a temperature induced  enhancement of the low-energy GT$_-$  strength followed by a remarkable quenching of 
  the high-energy component of the resonance was  observed in some closed-shell nuclei.

 \bibliography{library}


\end{document}